\documentclass[journal,comsoc]{IEEEtran}

\usepackage{threeparttable}
\usepackage{booktabs}
\usepackage{multirow}
\usepackage{makecell}

\usepackage[T1]{fontenc}
\usepackage{graphicx}
\usepackage{bm}
\usepackage[noend]{algpseudocode}
\usepackage{algorithmicx,algorithm}
\usepackage{threeparttable}
\usepackage[nocompress,sort]{cite}

\ifCLASSINFOpdf
\else
\fi
\usepackage{amsmath}
\usepackage[cmintegrals]{newtxmath}
\ifCLASSOPTIONcompsoc
  \usepackage[caption=false,font=normalsize,labelfont=sf,textfont=sf]{subfig}
\else
  \usepackage[caption=false,font=footnotesize]{subfig}
\fi
\usepackage{hyperref}
\hypersetup{hidelinks}

\def\equationautorefname#1#2\null{
	Eq. (#2\null)
}

\begin{document}
%
\title{Integrating Knowledge into End-to-End Speech Recognition from External Text-Only Data}
%
%
%

\author{Ye~Bai,~\IEEEmembership{Student Member,~IEEE,}
        Jiangyan~Yi,~\IEEEmembership{Member,~IEEE,}
        Jianhua~Tao,~\IEEEmembership{Senior Member,~IEEE}
        Zhengqi~Wen,~\IEEEmembership{Member,~IEEE}
        Zhengkun~Tian,~\IEEEmembership{Student Member,~IEEE}
        Shuai~Zhang,~\IEEEmembership{Student Member,~IEEE}
\thanks{This work is supported by the National Key Research and Development	Plan of China (No.2018YFB1005003), the National Natural Science Foundation of China (NSFC) (No.61831022, No.61901473, No.61771472, No.61773379 ) and an Inria-CAS Joint Research Project (No.173211KYSB20190049).}
\thanks{Ye Bai is currently a PhD candidate at University of Chinese Academy of Sciences (UCAS), Beijing, China. e-mail: baiye2016@ia.ac.cn}
\thanks{Jiangyan Yi is currently an Assistant Professor in NLPR, Institute of Automation, Chinese Academy of Sciences, Beijing, China. e-mail: jiangyan.yi@nlpr.ia.ac.cn.}
\thanks{Jianhua Tao is currently a Professor in NLPR, Institute of Automation,	Chinese Academy of Sciences, Beijing, China. e-mail: jhtao@nlpr.ia.ac.cn (Corresponding author: Jianhua Tao and Jiangyan Yi).}
\thanks{Zhengqi Wen is currently an Associate Professor in NLPR, Institute of Automation, Chinese Academy of Sciences, Beijing, China. e-mail: zqwen@nlpr.ia.ac.cn.}
\thanks{Zhengkun Tian is currently a PhD candidate at University of Chinese Academy of Sciences (UCAS), Beijing, China. e-mail: zhengkun.tian@nlpr.ia.ac.cn.}
\thanks{Shuai Zhang is currently a PhD candidate at University of Chinese Academy of Sciences (UCAS), Beijing, China. e-mail: shuai.zhang@nlpr.ia.ac.cn.}

}

%
%

\markboth{Journal of \LaTeX\ Class Files,~Vol.~14, No.~8, August~2015}%
{Shell \MakeLowercase{\textit{et al.}}: Bare Demo of IEEEtran.cls for IEEE Communications Society Journals}
%



\maketitle

\begin{abstract}
Attention-based encoder-decoder (AED) models have achieved promising performance in speech recognition. However, because of the end-to-end training, an AED model is usually trained with speech-text paired data. It is challenging to incorporate external text-only data into AED models. Another issue of the AED model is that it does not use the right context of a text token while predicting the token. To alleviate the above two issues, we propose a unified method called LST (Learn Spelling from Teachers) to integrate knowledge into an AED model from the external text-only data and leverage the whole context in a sentence. The method is divided into two stages. First, in the representation stage, a language model is trained on the text. It can be seen as that the knowledge in the text is compressed into the LM. Then, at the transferring stage, the knowledge is transferred to the AED model via teacher-student learning. To further use the whole context of the text sentence, we propose an LM called causal cloze completer (COR), which estimates the probability of a token, given both the left context and the right context of it. Therefore, with LST training, the AED model can leverage the whole context in the sentence. Different from fusion based methods, which use LM during decoding, the proposed method does not increase any extra complexity at the inference stage. We conduct experiments on two scales of public Chinese datasets AISHELL-1 and AISHELL-2. The experimental results demonstrate the effectiveness of leveraging external text-only data and the whole context in a sentence with our proposed method, compared with baseline hybrid systems and AED model based systems.
\end{abstract}


%
\IEEEpeerreviewmaketitle

\section{Introduction}
%
%
%
%


\IEEEPARstart{D}{eep} learning has carried automatic speech recognition (ASR) into real-world applications with its powerful representation ability. Conventionally, an ASR system consists of a signal processing module for feature extraction, an acoustic model (AM) based on hidden Markov models (HMMs) estimating probabilities of acoustic units (e.g. phone, syllable), a pronunciation lexicon mapping the acoustic units to words, and a language model (LM) estimating word sequence probabilities \cite{yu2016automatic}. The deep neural networks (DNNs) are used to estimate HMM observation probabilities. This DNN-HMM hybrid framework has achieved success due to advanced techniques, such as sequence discriminative training \cite{vesely2013sequence} and various network structures \cite{sainath2013deep,zeyer2017comprehensive,peddinti2015time}. However, in the DNN-HMM hybrid framework, the acoustic models and the language models are trained separately. It causes three issues. First, building a pronunciation lexicon needs expert knowledge which is difficult to obtain, especially in low-resource languages. Second, the static searching graph based on n-gram LMs trained on large-scale text data may be very large. Third, the separate training procedures of the AMs and the LMs cause error accumulation. 

Recently, pure neural network architectures, such as attention-based encoder-decoder (AED) models and RNN transducers, have achieved success in ASR \cite{chorowski2015attention,bahdanau2016endtoend,chan2016listen,kim2017joint,dong2018speech,chiu2018state,graves2012sequence,rao2017exploring}. The pure neural network architecture based systems model both acoustic and linguistic knowledge with neural networks. Different from the conventional hybrid models, these pure neural network architectures do not need pronunciation lexicons and static searching graphs. Moreover, the acoustic features and linguistic knowledge are modeled simultaneously, and the whole system is optimized in an end-to-end manner. These advantages draw the interests of the speech community.

However, because of the end-to-end training process, these pure neural network architectures need parallel speech-text data. It is often expensive to collect these parallel data. Therefore, the performance of the pure neural network ASR system is limited by the amount and the quality of the speech-text data. In contrast, the conventional hybrid ASR architectures use the large-scale text-only data to train LMs and improve the performance. The LMs of production-level hybrid ASR systems are often trained on dozens of gigabytes of text or more, and it makes the system generalize in open domains better. The rich linguistic knowledge in the large-scale text is useful and has been demonstrated the effectiveness to improve the performance in many natural language processing tasks \cite{peters2018deep,devlin2018bert}. Moreover, compared with the speech-text parallel data, the text-only data is cheap and convenient to collect. However, it is non-trivial to integrate the knowledge from the external text-only data into the end-to-end pure neural network based ASR systems.

Previous work \cite{gulcehre2015on,sriram2018cold,kannan2018an,shan2019component} have investigated fusion based methods to integrate an external LM into the end-to-end pure neural network system. In these work, the scores of the ASR model and the LM are fused with a weighting coefficient or a multi-layer perceptron (MLP). However, the external LM module increases complexity during inference. Therefore, it is worth to investigate a different method to integrate the knowledge from the text-only data without external modules at the inference stage.

Another issue of the decoder of an AED model is the lack of the use of the global text context in a sentence. Specifically, a common decoder of an AED model generates text in a left-to-right manner, i.e., it only uses the left context to predict the next word. Because of this autoregressive property, these decoders are difficult to leverage the right context, which influences the performance during decoding. This is because the right context is not used, a small mistake which happens at the left side causes more mistakes at the right side (which is referred to as exposure bias problem \cite{bengio2015scheduled}). This issue also exists in conventional speech recognition systems. 

Several previous work are proposed to use the right context for AED models. A forward-backward searching algorithm was proposed for an AED model to decode speech from left to right as well as right to left \cite{mimura2018forward}. However, the three-pass decoding algorithm increases complexity during inference. A synchronous bidirectional transformer model, which uses left-to-right and right-to-left decoding simultaneously and interactively for AED model based machine translation, does not need multi-pass decoding \cite{zhou2019synchronous}. However, the bidirectional attention computed in parallel makes the model complex. Bidirectional agreement methods, which minimizes discrepancy between a left-to-right decoding AED model and a right-to-left one, improves performance of AED models for machine translation \cite{liu-etal-2016-agreement-target,zhang2019regularizing} and end-to-end text-to-speech \cite{zheng2019forward}. However, in these methods, the agreement is made between the left-to-right model and the right-to-left model, but the whole sentence context is not leveraged in either of the models. Moreover, the left-to-right AED model and the right-to-left AED model are trained with the parallel speech-text data, which limits the flexibility of the methods.

Inspired by teacher-student learning \cite{2006model,hinton2015distilling,romero2014fitnets,li2014learning,li2017large}, which transfers knowledge from one model to another, we propose a unified training method to address the aforementioned two problems. The method consists of two steps.
\begin{enumerate}
	\item \textbf{Representation}. We use a neural network to represent knowledge of the external text. The LM predicts the probability of a word given the corresponding context. It can be seen as that the knowledge of the large-scale text is ``compressed'' into the neural network. 
	\item \textbf{Transferring}. We transfer the knowledge from the LM to the AED via teacher-student learning. The LM is used as the ``teacher'' model to provide soft labels of training transcriptions. And these soft labels are used to train the ``student'' AED model. Thus, the AED model can use the knowledge from the external text-only data.
\end{enumerate}
For the representation step, we propose a self-attention based language model called Causal clOze completeR (COR), which models the \textit{whole context} of a sentence. Therefore, with the above training method, the AED model not only learns from the external text-only data, but also from the whole context (including the left context and the right context). 

Because the proposed method can be seen as that the AED model learns ``spelling'' words from the ``teacher'' LM, we refer to the training method as ``Learn Spelling from Teachers'' (LST). The proposed method does not add any external module during inference. And the method is flexible to use non-parallel text rather than parallel speech-text data. Furthermore, the method can use the whole context of a sentence.

\begin{figure}[!t]
	\centering
	\includegraphics[width=1.0\columnwidth]{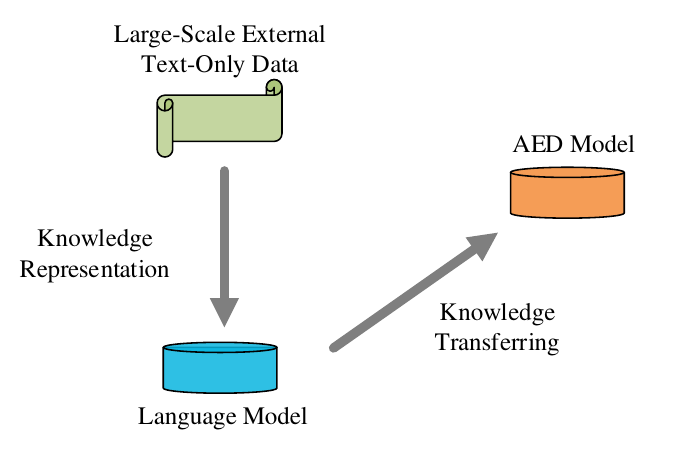}
	\caption{An illustration of the LST method. First, the knowledge in the external text-only data is represented as a language model. Then the knowledge is transferred to the attention-based encoder-decoder (AED) model via teacher-student learning.} 
	\label{fig:lst}
	\vspace{-15pt}
\end{figure}

The contributions of this work are summarized as follows.
\begin{itemize}
	\item \textit{A unified training method}. We present a unified training method to integrate knowledge from text into an AED model. First, knowledge in the text is represented with an LM. Then, the knowledge is transferred into the AED model.
	\item \textit{Whole context utilization}. We propose a new self-attention mechanism based LM COR, which utilizes both the left context and the right context. Compared with autoregressive LMs, COR leverages the whole context rather than the left context. By transferring the knowledge from COR, the AED model can use the knowledge of the whole context in a sentence.
	\item \textit{No extra module during inference}. The all procedure is at the training stage. The proposed methods do not increase complexity during inference.
\end{itemize}

This journal version paper is extended from our conference paper of INTERSPEECH 2019 \cite{bai2019learn}. The new content in this paper includes a new LM which represents whole sentence knowledge, a systematic description of the proposed method, a detailed analysis of the behaviors during inference, more detailed experiments on two scales of datasets, and better performance of the systems. The rest of this paper is organized as follows. \autoref{sec:gb} introduces the background AED models. \autoref{sec:lst} describes the proposed LST training method. \autoref{sec:cor} introduces the proposed COR LM. \autoref{sec:rel} reviews and compares the related work. \autoref{sec:exp} describes the experiments and analysis. At last, \autoref{sec:conc} summarizes the paper and introduces the future work.

\section{Background: Attention-Based Encoder-Decoder Models}
\label{sec:gb}
\begin{figure}[!t]
	\centering
	\includegraphics[width=1.0\columnwidth]{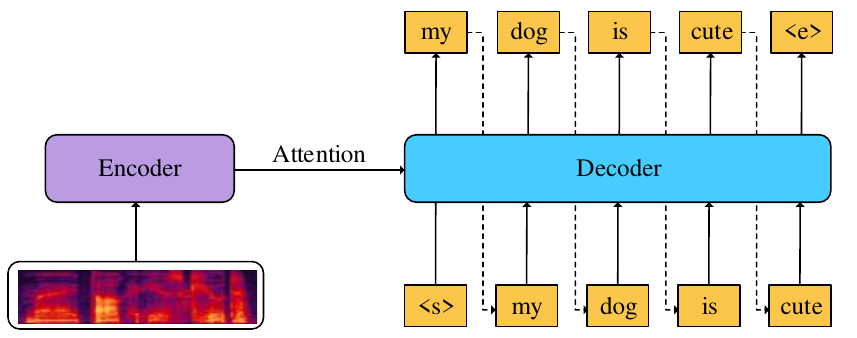}
	\caption{The architecture of the attention-based encoder-decoder model. \texttt{<s>} denotes the start-of-sentence token, and \texttt{<e>} denotes the end-of-sentence token. } 
	\label{fig:las}
\end{figure}

In this section, we briefly review the AED model. As shown in \autoref{fig:las}, an AED model consists of three parts. The encoder extracts high-level representations from speech. The decoder generates the text sequence in terms of the high-level representations. The attention mechanism bridges the encoder and the decoder. It uses the hidden state of the decoder to query the most matching high-level acoustic representation from the encoder outputs, and the decoder predicts the next word with the previous words and the queried acoustic representation. In this case, the encoder ``listens'' to the speech. The decoder ``attends'', and ``spells'' the words. Thus, this framework is also referred to as ``Listen, Attend, and Spell'' (LAS) \cite{chan2016listen}.

Formally, given a speech-text pair $(X, Y)$, where $X=[x_1, \cdots, x_I]$ denotes the acoustic feature sequence, and $Y=[y_1, \cdots, y_J]$ denotes the text token (word, sub-word, or phone) sequence, the AED model estimates the conditional probability:
\begin{equation}
	\label{eq:arm}
	P_{AED}(Y|X) = P(y_1|X)\prod_{j=2}^{J} P(y_j|Y_{<j}, X).
\end{equation}
$x_i$ denotes $i$-th acoustic feature in the sequence. $y_j$ denotes $j$-th token in the token sequence. $I$ and $J$ are the lengths of the two sequences. $Y_{<j}$ denotes the subsequence $[y_1, \cdots, y_{j-1}]$. The first token $y_1$ is always \texttt{<s>} which denotes the start of the sentence, so $P(y_1|X) = 1$. And $y_J$ is \texttt{<e>} which denotes the end of the sentence.

The encoder encodes the acoustic feature sequence $X$ into high-level representations $Z$:
\begin{equation}
\label{eq:enc}
	Z = \text{Encode}(X),
\end{equation}
where $Z=[z_1, \cdots, z_{I_{\text{sub}}}]$ is the output sequence of the encoder. Length $I_{\text{sub}}$ is usually smaller than length $T$ due to subsampling mechanism. The decoder estimates the probability distribution over the token vocabulary with the attention mechanism, given $Z$ and the previous tokens $Y_{<j}$:
\begin{equation}
\label{eq:dec}
P(y_j|Y_{<j}, X) = \text{AttendAndDecode}(Y_{<j}, Z).
\end{equation}
For the recurrent neural network (RNN) based decoder, the previous tokens $Y_{<j}$ are encoded as some latent context vector. For the recent attention-based non-recurrent feedforward network \cite{vaswani2017attention}, the previous tokens $Y_{<j}$ are inputted into the decoder together.

The parameters of the AED model are trained with cross-entropy loss:
\begin{equation}
\label{eq:ce}
L_{CE}(\theta) = -\frac{1}{N} \sum_{n=1}^{N} \frac{1}{J^{(n)}-1} \sum_{j=2}^{J^{(n)}} \log P_{AED}(y_j^{(n)}|Y_{<j}^{(n)}, X^{(n)}; \theta).
\end{equation}
where $(X^{(n)}, Y^{(n)})$ is the $n$-th pair of data in the corpus. $J^{(n)}$ is the corresponding length. $N$ is the total number of the data. $\theta$ represents the parameters of the whole neural network. 

\begin{algorithm*}[!t]
	\caption{The LST training method.} 
	\hspace*{0.02in} {\bf Input:} 
	A speech-text paired dataset and a pretrained LM. Hyper parameters $\lambda$ and $T$.  \\
	\hspace*{0.02in} {\bf Output:} 
	The parameters of the AED model.
	\begin{algorithmic}[1]
		\State Initialize the parameters of the AED model randomly. 
		\While{ not converged }{} 
		\State Randomly select a minibatch of data $\{(X^{(n)}, Y^{(n)})\}_{n=b}^{b+B}$ with batch size $B$.
		\State Input acoustic feature sequences $X^{(n)}$ into the AED model, and forward propagate.
		\State Input transcriptions $Y^{(n)}$ into the LM, and forward propagate. Obtain probability distributions $P_{LM}(y_j|C^{(n)})$ with temperature $T$ in \autoref{eq:lm-softmax}.
		\State Compute cross-entropy loss $L_{CE}(\theta)$ in \autoref{eq:ce}.
		\State Compute LST loss $L_{LST}(\theta)$ in \autoref{eq:lstloss}.
		\State Combine the two losses with $\lambda$ in \autoref{eq:loss}
		\State Back propagate the loss and update the parameters $\theta$.
		\EndWhile
		\Return $\theta$
	\end{algorithmic}
	\label{alg:update}
\end{algorithm*}

\section{Learn Spelling from Teachers}
\label{sec:lst}

\begin{figure}[!t]
	\centering
	\includegraphics[width=1.0\columnwidth]{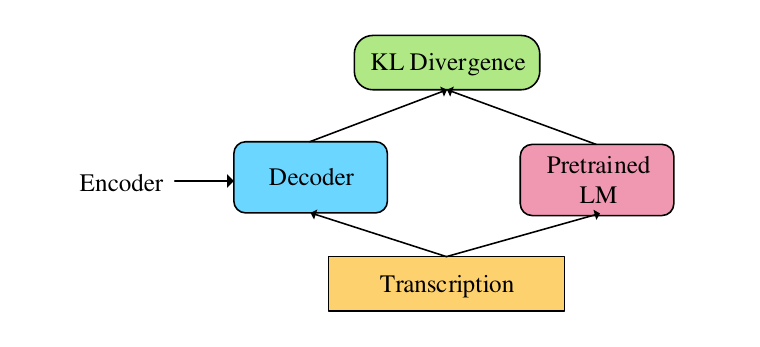}
	\caption{The computation of LST loss.} 
	\label{fig:lstloss}
\end{figure}

In this section, we introduce the proposed LST training method. As shown in \autoref{fig:lst}, the knowledge in the large-scale external text-only data is represented as an LM. Then, the knowledge is transferred to the AED model.

Specifically, the LM which is trained on external text-only data provides probability distribution corresponding to each token in the transcription. And the Kullback-Leibler divergence (KLD) between the decoder of the AED model and the LM is minimized. 

Here, the LM is a general model to estimate the probability of a token in the sentence, given the context. This makes the method more flexible. We denote the probability of $j$-th token in the transcription given by the LM as $P_{LM}(y_j|C)$, where $C$ represents the context. 

The KLD between the decoder of the AED model and the LM for $j$-th token is:
\begin{equation}
\label{eq:kld}
D(P_{LM} || P_{AED} ) = \sum_{y_j \in \mathcal{S}} P_{LM}(y_j|C) \log \frac{P_{LM}(y_j|C)}{P_{AED}(y_j|Y_{<j}, X; \theta)},
\end{equation}
where $\mathcal{S}$ is the vocabulary with size $M$. Because the parameters of the LM are not updated when we train the AED model, the KLD can be simplified to cross-entropy. The LST loss is defined as:
\begin{equation}
\label{eq:lstloss}
\begin{split}
&  H_j^{(n)}(\theta) = -\sum_{y_j \in \mathcal{S}} P_{LM}(y_j|C^{(n)})\log P_{AED}(y_j|Y_{<j}^{(n)}, X^{(n)}; \theta), \\
&L_{LST}(\theta) = -\frac{1}{N} \sum_{n=1}^{N} \frac{1}{J^{(n)}-1} \sum_{j=2}^{J^{(n)}} H_j^{(n)}(\theta),
\end{split}
\end{equation}
where $H_j^{(n)}$ is the cross-entropy for the token at $j$-th position in $n$-th sample of the corpus, and $C^{(n)}$ is a part of the transcription in $n$-th sample as the context for the LM. For an autoregressive LM, $C^{(n)}$ is the $Y_{<j}$, i.e. the left context. In \autoref{sec:cor}, we introduce the COR model which leverages the left context and the right context at the same time. 

To combine the knowledge of the groundtruth of the transcription and the knowledge from the LM, $L_{CE}(\theta)$ and $L_{LST}(\theta)$ are added together:
\begin{equation}
\label{eq:loss}
L(\theta) = (1-\lambda)L_{CE}(\theta) + \lambda L_{LST}(\theta),
\end{equation}
where $\lambda \in [0,1]$ is a weight to balance $L_{CE}$ and $L_{LST}$.

Following \cite{hinton2015distilling}, we use a temperature parameter $T$ to smooth the probability of the LM $P_{LM}$. $P_{LM}(y_j|C)$ is usually computed by a softmax function:
\begin{equation}
\label{eq:lm-softmax}
o_m = \frac{ \exp(r_m / T )   }{ \sum_{m=1}^{M} \exp(r_m / T) },
\end{equation}
where $o_m$ is the probability of the $m$-th token in the vocabulary. $M$ is the size of the vocabulary. $r_m$ denotes the $m$-th node of the output. With larger temperature $T$, the probability distributions of the LM are smoother. Note that this is for neural network based LMs, which are used in this paper. For n-gram LMs, other smoothing methods can be used.

The AED model is trained with error propagation by stochastic gradient descent algorithm. The update procedure of the parameters is shown in \autoref{alg:update}.

\begin{figure}[!t]
	\centering
	\includegraphics[width=0.8\columnwidth]{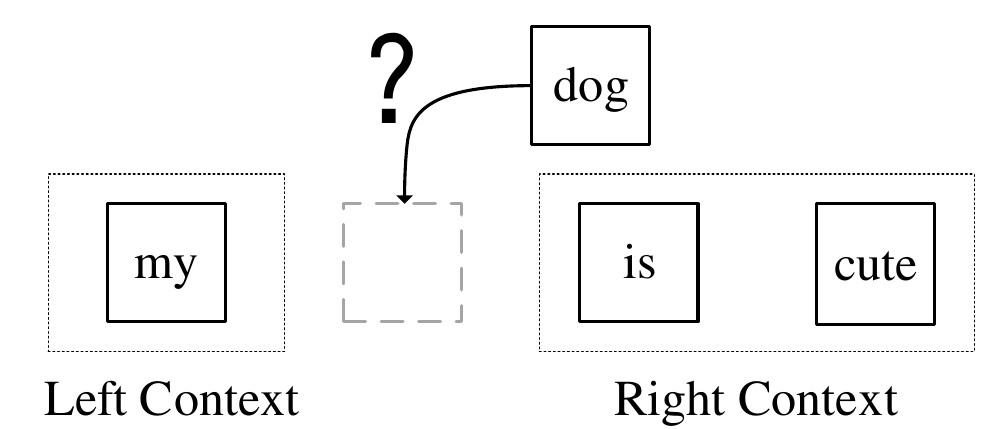}
	\caption{An example of a cloze test. We use the left context (``my'') and the right context (``is'' and ``cute'') to fill in the blank. That is, we want to model the probability $P(\text{dog}| \text{my}, \text{is}, \text{cute})$.} 
	\label{fig:cloze_example}
\end{figure}

\begin{figure*}[!t]\centering
	\subfloat[Causal Cloze Completer] 
	{ \label{fig:clozer_arch}
		\includegraphics[width=1.0\columnwidth]{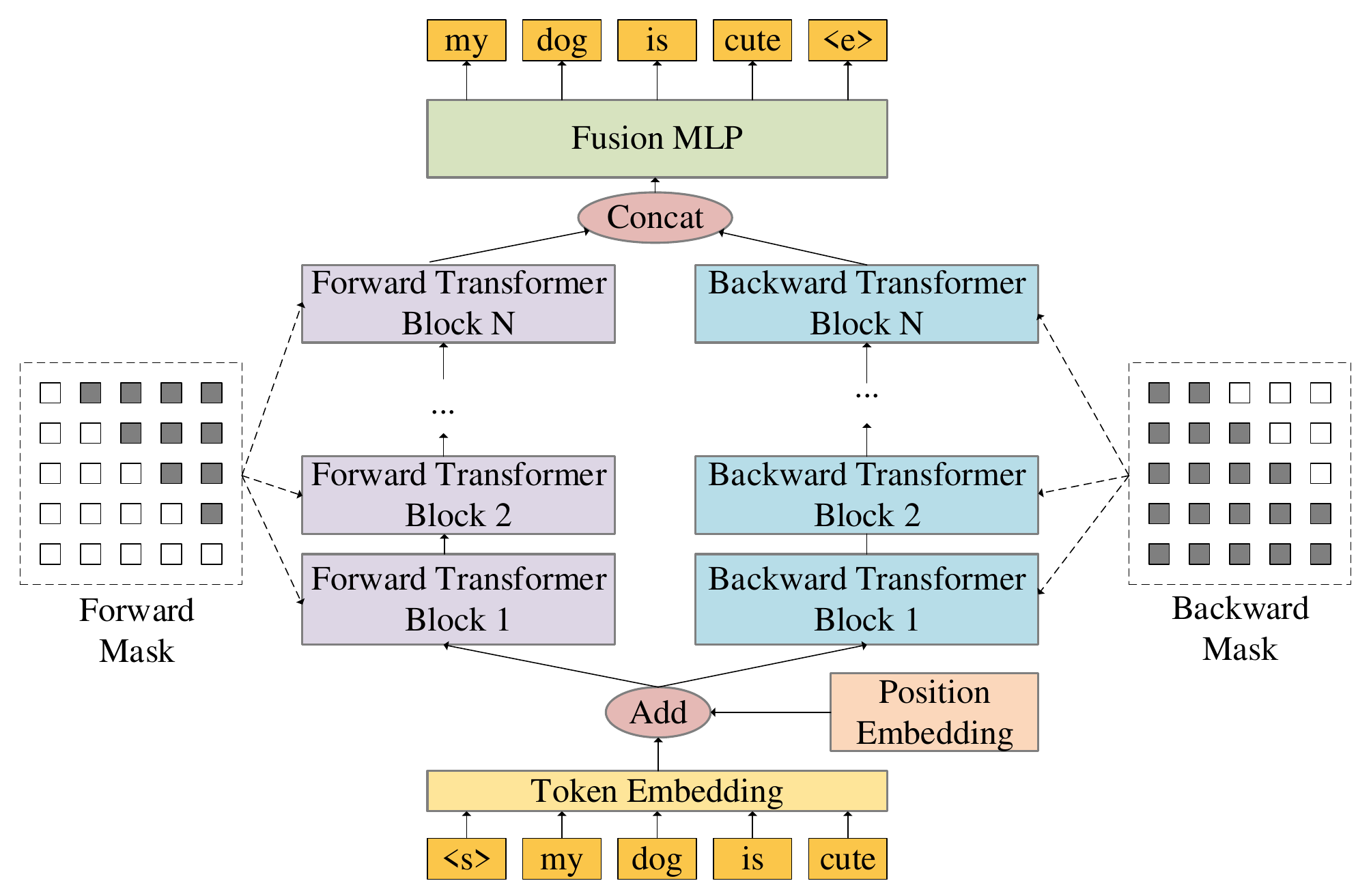}
	}		
	\qquad
	\subfloat[Transformer Block]  
	{ \label{fig:block}
		\includegraphics[width=0.3\columnwidth]{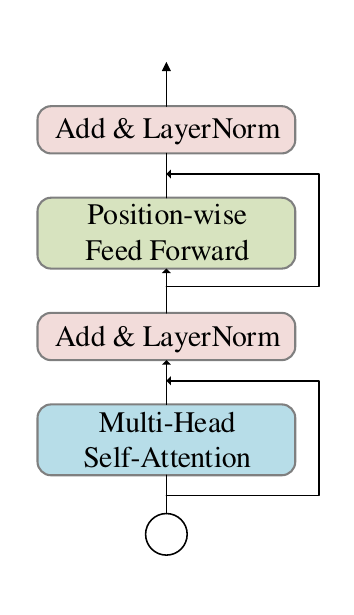}
	}
	\qquad
	\subfloat[Context Flow]  
	{ \label{fig:flow}
		\includegraphics[width=0.3\columnwidth]{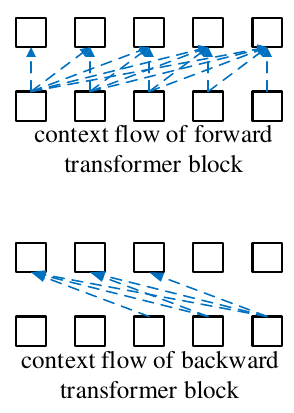}
	}  
	\label{fig:cor}
	\caption{(a) illustrates the architecture of COR. The input token sequence is first mapped to the corresponding embeddings and position embeddings are added. Then, the embeddings are inputted into the forward transformer stack and the backward transformer stack in parallel. At last, after a multi-layer perceptron, the probability distributions are computed for each position by a softmax function. The gray squares in mask matrices are set to zero. (b) illustrates the structure of a transformer block. (c) illustrates the context flows of the transformer blocks.} 
\end{figure*}

\section{Integrating whole context of a sentence}
\label{sec:cor}
In this section, we propose an attention-based feedforward LM called Causal clOze completeR (COR) for integrating the left context and the right context simultaneously. First, we introduce the cloze test problem. Then, we present the COR model.

\subsection{Cloze Test}
Motivated by previous pre-training language model work BERT \cite{devlin2018bert}, we introduce cloze test problem \cite{taylor1953cloze}. \autoref{fig:cloze_example} shows an example. In the cloze test, we predict a token in the sentence in terms of the left context and right context. In this procedure, the left context and right context, i.e., the whole sequence context, are used simultaneously.

The above procedure can be formulated as follows. Let a token sequence be $ Y = [y_1, \dots, y_J] $, where each $y_j$ represents a token in vocabulary, and $J$ is the sequence length. For the token $y_j$, given the left context $[y_1, \cdots, y_{j-1}]$ and the right context $[y_{j+1}, \cdots, y_{J}]$, we would like to predict the probability $P(y_j | y_1, \cdots, y_{j-1}, y_{j+1}, \cdots, y_{J})$.

\subsection{Causal Cloze Completer}

We use COR, a neural network, to estimate $P(y_j | y_1, \cdots, y_{j-1}, y_{j+1}, \cdots, y_{J})$ for each step in parallel. The self-attention mechanism is used to model the token sequence and predict the probabilities. The self-attention mechanism directly models the long-term dependency between each token in a sequence without recurrent structure, and it can be computed in parallel. 

The main problem of using the whole sentence context is how to prevent the model from ``seeing itself'' when it predicts a token. It makes the prediction trivial. For BERT \cite{devlin2018bert}, \texttt{[MASK]} token is used to randomly replace some tokens, and the model predicts these masked tokens during training. However, this does not match our downstream LST task, i.e. we need to achieve the probability distributions over the vocabulary at each position. For instance, for the sentence ``my dog is cute'', if we directly input the whole sentence into BERT, we achieve $P(my | my, dog, is, cute)$ at the first position. This is non-sense. A solution to this problem is that we can replace each token with \texttt{[MASK]} and compute each probability one-by-one, however, this needs multi-pass propagation for one sentence. 

Different from BERT \cite{devlin2018bert}, we do not introduce \texttt{[MASK]} token into the input sequence, but directly model the conditional probability $P(y_j | y_1, \cdots, y_{j-1}, y_{j+1}, \cdots, y_{J})$. This avoids the aforementioned issue, because we use attention masks to control the context flow rather than the \texttt{[MASK]} token. Specifically, we mask out some attention scores to zero to control the context flow and make the representation of one token itself cannot be passed to the next layer, as shown in \autoref{fig:flow}. 

Another problem is that the AED model is autoregressive, so there is an offset between the input and the output of the decoder. Previous bidirectional LMs \cite{mousa2017contextual,peters2018deep,devlin2018bert,baevski2019cloze} are autoencoder structure, i.e., the position of the input and the output are the same. This does not match the autoregressive decoder of the AED. To imitate the causality of the autoregressive AED model, we add an offset between the inputs and the outputs.

The architecture of COR is shown in \autoref{fig:clozer_arch}. It consists of two stacks of transformer blocks. The one uses the left context, and the other uses the right context. Then, the left context and the right context are fused to predict the probabilities with a fusion layer.

The main structure of a transformer block \cite{vaswani2017attention} is shown in \autoref{fig:block}. First, the inputs $U \in \mathbb{R}^{J \times D}$ with length $J$ are  aggregated with the following multi-head attention:
\begin{equation}
\label{eq:ori_att}
\begin{split}
&\text{Atten}(Q, K, V) = \text{Softmax}(\frac{QK^T}{\sqrt{D_{k}}} + M_A)V,  \\ 
&head_h = \text{Atten}(UW_h^q, UW_h^k, UW_h^v), h = 1,\cdots,N_H, \\
&\text{MHA}(U) = \text{Concat}(head_1,\cdots,head_{N_H})W^o, \\
\end{split}
\end{equation}
$Q \in \mathbb{R}^{T_q \times D_k} $, $K \in \mathbb{R}^{T_k \times D_k}$, $V \in \mathbb{R}^{T_k \times D_v}$ denote queries, keys, and values in the scaled dot-product attention, respectively. $T_q$ and $T_k$ are input lengths. $D_k$ and $D_v$ are corresponding dimensionality. $M_A$ is a mask matrix to control the context flow. The multi-head attention projects the input into different subspaces with parameter matrices $W_h^q$, $W_h^k$, $W_h^v$, and then concatenates the attended scores together. Here, the queries, keys, and values are all the input $U$, i.e., self-attention. So, $T_k=T_q=J$, and $D_k=D_v=D$. $N_H$ is the number of the heads in the multi-head attention.

\begin{figure}[!t]\centering
	\includegraphics[width=0.9\columnwidth]{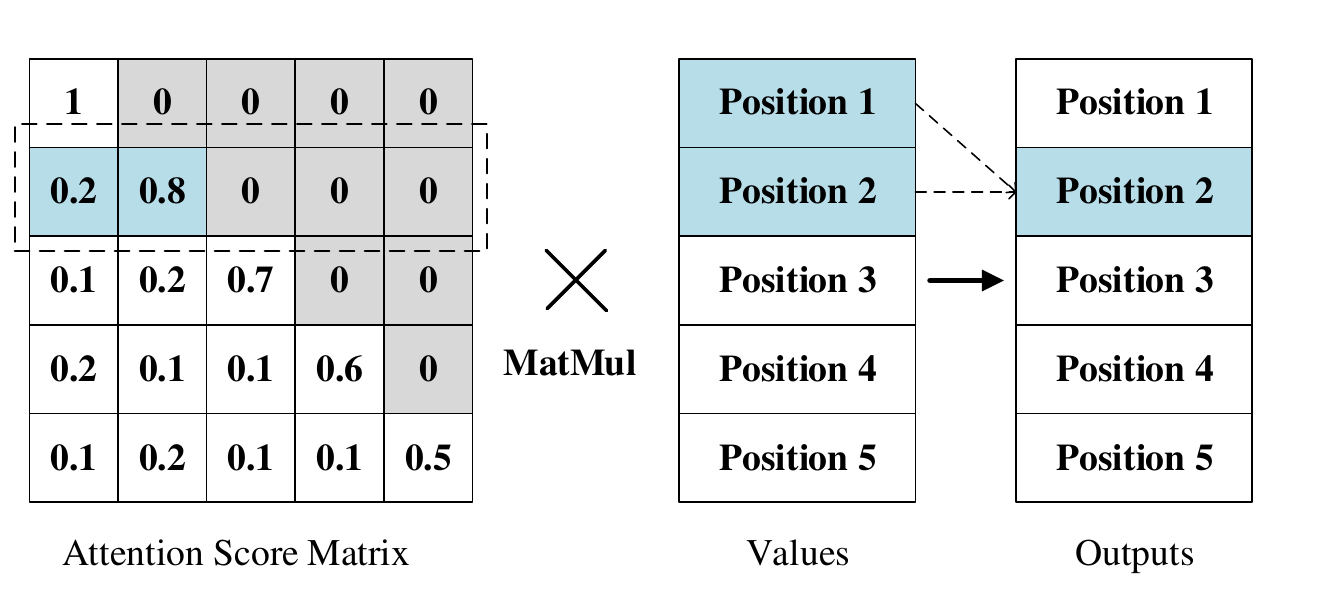}
	\caption{An illustration of using mask to control context flow. The attention score matrix is generated after the softmax function, and is used to fuse the input sequence (values). The masked scores can prevent the corresponding vector from being fused.} 
	\label{fig:mask}
\end{figure}

$M_A \in \{0, -\infty\}^{J \times J}$ is as follows:
\begin{equation}
\label{eq:mask}
M_A = \left\{
\begin{array}{rcl}
-\infty,          & & \text{masked out} \\
0.                 & & \text{otherwise}
\end{array} \right.
\end{equation}
With matrix $M_A$, the attention scores are masked to zero, and this part of the inputs is not used. \autoref{fig:mask} shows how the masks prevent the vectors in the inputted sequence being propagated to the outputs. 

After attention, a position-wise feed-forward network (FFN) is used as a non-linear transformation.
\begin{equation}
\label{eq:nonlinear}
s = \text{Activate}(e W_1 + b_1)W_2 + b_2.
\end{equation}
where $e$ is an input of FFN in the sequence, $s$ is an output of FFN in the sequence, $W_1$, $b_1$, $W_2$, and $b_2$ are parameters, and $\text{Activate}$ is the activation function. In this work, we use GLU \cite{dauphin2017language} as the activation function. The residual connections \cite{he2016} and layer normalization \cite{ba2016} are also used.  

The illustration of COR is shown in \autoref{fig:clozer_arch}. First, the input token sequence is embedded and added with position encodings. We use sinusoidal position encodings \cite{vaswani2017attention}. Then, two stacks of transformer blocks are used in parallel. The one is a stack of forward transformer blocks to model the left context, and the other is a stack of backward transformer blocks to model the right context. The outputs of the top forward transformer block and the backward transformer block are concatenated together and inputted a fusion multi-layer perceptron. The probability of each token at each step is computed with a softmax function. To match the decoder of the AED model, the input sequence of COR is from step $1$ to step $J-1$, and the output sequence of COR is from step $2$ to step $J$. Because we imitate the causal property, we name the model as causal cloze completer. 

The context flows are controlled by the $M_A$. The illustration of the forward mask matrix and the backward matrix is shown in \autoref{fig:clozer_arch}. For the forward transformer block, the output at step $j$ only ``sees'' the left context, i.e. the attention scores from $j$ to $J-1$ are masked to zeros. For the backward transformer block, the output at step $j$ only ``sees'' the right context, i.e. the attention scores from $1$ to $j-1$ are masked to zeros. Note that the output sequence has an offset to the input sequence, so the attention scores also have an offset. \autoref{fig:flow} shows the context flow of the blocks. For backward masks, some all-zero-row scores exist, and it makes softmax function illegal. In this case, we mask these rows to zeros after softmax.

The model is trained with cross-entropy loss. After optimization, COR predicts the probability distribution on the vocabulary at each step, given the left context and the right context.
\begin{equation}
P_{COR}(y_j | y_1, \cdots, y_{j-1}, y_{j+1}, \cdots, y_{J-1}) = \text{COR}(Y). j=2,\cdots,J
\end{equation}

\section{Related Work}
\label{sec:rel}
In this section, we review and compare previous work related to our work.

\vspace{-10pt}
\subsection{Knowledge Distillation}
Knowledge distillation (KD) was first proposed for model compression \cite{2006model,li2014learning,hinton2015distilling}. In KD, a large but powerful model is trained on training data. Then, a small-size model is trained to mimic the large model and to achieve close performance. Because this mimic property, it is also referred to as teacher-student learning. Yoon et al. proposed a sequence-level KD to reduce AED model size for machine translation \cite{kim2016sequence}. KD based method is extended for domain adaptation in acoustic modeling \cite{li2017large} and language modeling \cite{andres2018efficient}. Different from previous work, this paper focuses on how to transfer knowledge from the external text-only data to AED models.

\vspace{-10pt}
\subsection{Label Smoothing}
Label smoothing is used as an effective regularization method to prevent DNNs from over-confidence problems \cite{szegedy2016rethinking,pereyra2017regularizing}. It discounts the original one-hot label with a small factor and then allocates the factor to all classes uniformly. Label smoothing has become a standard method to improve the performance of AED models \cite{chorowski2017towards}. It can be seen as a special case of teacher-student learning when the teacher is uniform distribution \cite{pereyra2017regularizing}. A variant of label smoothing is distributing the factor to the classes in terms of class frequency, which is referred to as unigram smoothing. When training AED models for ASR, label smoothing can also be seen as a special case of LST. Specifically, uniform smoothing can be seen that the teacher LM is zero-gram LM, and unigram smoothing can be seen that the teacher LM is unigram LM. Different from previous work, we use neural network to provide a context-dependent prior distribution rather than a simple uniform distribution. The prior distribution is estimated in a data-driven manner without any assumption.

\vspace{-10pt}
\subsection{Bidirectional Agreement}
Several previous work optimize KLD between a left-to-right AED model and a right-to-left one \cite{zheng2019forward,liu-etal-2016-agreement-target,zhang2019regularizing} to leverage the right context and improve the model performance. In these work, a left-to-right AED model and a right-to-left one are trained in advance, and then the right-to-left model provides soft labels to optimize the left-to-right one. However, these methods require parallel data which limits flexibility. Different from these work, we train COR which only leverages external text-only data and can model left context and right context simultaneously. So the methods are more flexible.

\vspace{-10pt}
\subsection{Whole-Sentence Language Models}
Maximum entropy based language models are used model whole-sentence probability with exponential functions \cite{rosenfeld2001whole,chen2009shrinking,amaya2001improvement}. Trans-dimensional random fields are also used to model whole-sentence probabilities\cite{wang2017learning}. These work are often used to rescore ASR hypotheses. However, these methods need complex sampling methods to optimize the language models. And probabilities at each token position are not computed with these models. Bidirectional recurrent neural networks (Bi-RNNs) are used as language models \cite{arisoy2015bi,he2016on,chen2017investigating}. However, these models only compute pseudo-likelihoods rather than true likelihoods \cite{arisoy2015bi}. These models are also used in ASR rescoring. Recently, whole-sentence pre-trained language models improved downstream tasks significantly \cite{devlin2018bert}. Bi-RNNs \cite{mousa2017contextual,peters2018deep} and transformer networks \cite{devlin2018bert,zhang2019ernie,Yang2019XLNet} are used as language models and improve downstream natural language processing tasks. COR also leverages transformers. Different from previous BERT-like work, we do not use \texttt{[MASK]} token but directly model the two-side context, which reduces the mismatch between our downstream task LST and the inference of BERT. Previous work \cite{baevski2019cloze} also uses two stacks of transformers. However, their network is not designed for text generation, so their network is not causal. Our goal is to transfer the whole-context knowledge to the AED model, so we imitate the causal property, i.e. there is an offset between the inputs and the outputs. 

\section{Experiments}
\label{sec:exp}
In this section, we introduce the experiments to show the effectiveness of the proposed method. First, we introduce the datasets used in this paper. Second, we introduce the experimental setup. Then, we introduce the experimental results and analysis.

\begin{table}[!t]\centering
	\caption{The structure of AISHELL-1 dataset.}
	\begin{tabular}{lccc}
		\toprule
		& \#utter. & \#hours & \#speaker \\ \midrule
		Training    & 120,098 & 150     &   340        \\
		Dev. & 14,326  & 18      & 40          \\
		Test        & 7,176   &  10     &  20         \\ \bottomrule
	\end{tabular}
	\label{tab:aishell1}
\end{table}

\begin{table}[!t]\centering
	\label{tab:aishell2}
	\caption{The structure of AISHELL-2 dataset.}
	\begin{tabular}{lccc}
		\toprule
		& \#utter.   & \#hours  & \#speaker \\ \midrule
		Training (iPhone)  & 1,009,223 & 1000     &   1347        \\
		Dev. (iPhone)      & 2,500     & 2        & 5          \\
		Dev. (Android)     & 2,500     & 2        & 5          \\
		Dev. (HiFi Mic.)   & 2,500     & 2        & 5          \\
		Test (iPhone)      & 5,000     &  4       &  10         \\
		Test (Android)     & 5,000     &  4       &  10         \\ 
		Test (HiFi Mic.)   & 5,000     &  4       &  10         \\  \bottomrule
	\end{tabular}
\end{table}

\subsection{Datasets}

\textbf{Speech datasets}. We conducted experiments on two public Chinese speech datasets AISHELL-1 \cite{bu2017aishell} and AISHELL-2 \cite{du2018aishell} to demonstrate the effectiveness in different scales of data. The detailed information of the datasets is as follows.
\begin{enumerate}
	\item AISHELL-1\footnote{http://www.openslr.org/33/} contains 178 hours of Mandarin speech, which is recorded with a high fidelity microphone in 44.1 kHz. Then, all audio is re-sampled to 16 kHz and stored in a 16-bit PCM format. The speech is recorded by 400 speakers, which are balanced with their gender, ages, and birth-place. The text transcriptions cover 5 domains, i.e., ``Finance'', ``Science and Technology'', ``Sports'', ``Entertainments'', and ``News''. The dataset is divided into three parts. The training set contains about 150 hours of speech. The development set contains about 18 hours of speech. And the test set contains about 10 hours of speech.
	\item AISHELL-2\footnote{http://www.aishelltech.com/aishell\_2} contains about 1000 hours of Mandarin speech. All audio is stored in 16 kHz, 16-bit PCM format. The topics of the transcriptions cover voice commands, digital sequence, places of interest, entertainment, finance, technology, sports, English spellings, and free speaking without specific topics. The training set is about 1000 hours of speech recorded by 1991 speakers iPhone smartphones. The development set contains about 2 hours of speech recorded by 5 speakers. And the test set contains about 4 hours of speech recorded by 10 speakers. Both development set and test set are recorded with the three types of equipment in parallel, i.e., iPhone smartphones, Android smartphones, and high fidelity microphones (HiFi Mic.).
\end{enumerate}

\textbf{External text}. We extract subsets of CLMAD \cite{yebai2018clmad,li2007scalable} text dataset as external text for training LMs. An open source tool XenC to extract the subset of CLMAD which is topic matched \cite{rousseau2013xenc}. The preprocessing steps are as follows:
\begin{enumerate}[]
	\item Select $3$ million sentences which have small cross-entropy transcriptions \cite{moore2010intelligent}; 
	\item Remove the sentences which are too long;
	\item Mix the remained sentences with training transcriptions;
	\item Segment the word sequences into characters.
\end{enumerate}
The description of the text data for AISHELL-1 and AISHELL-2 are shown in \autoref{tab:text1}. We mix $10$ times of training transcriptions with the selected subset of the external text for AISHELL-1. And $5$ times for AISHELL-2. This makes the external text more matching\footnote{We have released the preprocessed text data. The external text for AISHELL-1 can be downloaded at https://1drv.ms/u/s!An08U7hvUohBb234-V-Z0Qb\_Zcc?e=fK02E0, and the external text for AISHELL-2 can be downloaded at https://1drv.ms/u/s!An08U7hvUohBcznPgD5Io0AZlrU?e=UUhkDF.}.

Domain topic influences the performance of LMs. To show that the selected external text is topic domain matched, we train trigram LMs on training transcriptions and the external text, respectively, and we compute the perplexities on the development set. The results are shown in \autoref{tab:ppl1} and \autoref{tab:ppl2}. We can see that the trigram LM trained on the external text achieves much lower perplexity (PPL) on the development set, compared with training transcriptions. This demonstrates that the external text is topic matched, and can improve the performance of LM.

\begin{table}[!t]	
	\caption{The description of the text for AISHELL-1. }
	\centering
	\begin{tabular}{ l  r  r  r } 
		\toprule
		& \#Sentences    &    \#Characters  &    Size              \\
		\midrule
		Training Trans.  & $120,098$      &    $1,730,113$   &    $5.1$MB           \\
		Dev. Trans.  	 & $14,326$       &    $205,341$     &    $0.8$MB           \\		
		Test Trans.      & $7,176$        &    $104,765$     &    $0.4$MB           \\
		External Text    & $3,703,982$    &    $75,893,998$  &    $221$MB            \\
		\bottomrule
	\end{tabular}
	\label{tab:text1}		
\end{table}

\begin{table}[!t]	
	\caption{The description of the text for AISHELL-2. }
	\centering
	\begin{threeparttable}
		{	
			\begin{tabular}{ l  r  r  r } 
				\toprule
				& \#Sentences     &    \#Characters   &    Size               \\
				\midrule
				Training Trans.  & $1,009,223$     &    $10,995,287$   &    $33$MB             \\
				Dev. Trans. * 	 & $2,500$         &    $24,802$       &    $0.08$MB           \\		
				Test Trans. *      & $5,000$         &   $104,765$      &   $0.16$MB           \\
				External Text    & $7,874,474$     &    $130,895,577$   &   $381$MB            \\
				\bottomrule
		\end{tabular}}
		\begin{tablenotes}	
			\footnotesize
			\item[*] For AISHELL-2, the content recorded with different equipment are the same. Thus, we only list one of them. 
		\end{tablenotes}
	\end{threeparttable}
	\label{tab:text2}	
\end{table}

\subsection{Experimental Setup}
\textbf{AED models}. We use Speech-Transformer \cite{dong2018speech} as the AED model. Speech-Transfermer is a non-recurrent AED model which uses self-attention mechanisms \cite{chorowski2015attention} to captures long-term dependency. Both encoder and decoder have $6$ transformer blocks. The dimensionality of the model is $512$, the number of heads is $8$, and the dimensionality of the position-wise feed-forward module is $2048$. The activation functions are GLUs. We use a two-layer convolutional neural network (CNN) for subsampling the inputted acoustic feature sequence. Each layer has $32$ convolution filters with a size $3\times3$, and the stride on the time axis is $2$. Thus,the frame rate is subsampled to $1/4$. The activation functions of the CNN are ReLUs.

We use $80$-dimension Mel-filter bank features (FBANK) as the inputs, which are extracted every 10ms with 25ms of frame length. The token vocabulary is composed of all characters in the training set and three special tokens. ``\texttt{<unk>}" token denotes the unknown token, ``\texttt{<s>}" represents the start of a sentence, and ``\texttt{<e>}" represents the end of a sentence.  The vocabulary size for AISHELL-1 is $4232$, and the vocabulary size for AISHELL-2 is $5252$.

\begin{table}[!t]	
	\caption{Perplexities of trigram LMs trained with training transcriptions and external text on the development set of AISHELL-1.   }
	\centering
	\begin{tabular}{ l  c} 
		\toprule
		\multicolumn{1}{c}{LM}  & PPL on dev. set   \\
		\midrule
		Training Trans.  & $70$         \\
		External Text    & $47$       \\
		\bottomrule
	\end{tabular}
	\label{tab:ppl1}		
\end{table}

\begin{table}[!t]	
	\caption{Perplexities of trigram LMs trained with training transcriptions and external text on the development set of AISHELL-2. }
	\centering
	\begin{tabular}{ l  c} 
		\toprule
		\multicolumn{1}{c}{LM}	& PPL on dev. set   \\
		\midrule
		Training Trans.  & $78$         \\
		External Text    & $62$       \\
		\bottomrule
	\end{tabular}
	\label{tab:ppl2}		
\end{table}

\begin{table*}[!t]	
	\caption{AISHELL-1: The selection of hyperparameters $\lambda$ and $T$ of LST training on the development set for each LM. The value in bold font denotes the selected hyperparameters. }
	\label{tab:aishell1-hyper}
	\centering
	\begin{tabular}{cccccccc}
		\toprule
		\multirow{2}{*}{$\lambda$} & \multirow{2}{*}{$T$} & \multicolumn{6}{c}{CER\%}                                                               \\ \cmidrule(l){3-8} 
		&  & LSTM LM (Trans.) & LSTM LM (Ext.) & Transformer LM (Trans.) & Transformer LM (Ext.) & COR (Trans.) & COR (Ext.) \\ \midrule
		0.1                & 1.0                & 9.2          & 9.2          & 10.4         & 8.9          & 8.7          & 8.7          \\
		0.1                & 2.0                & 7.8          & 8.1          & 7.9          & 7.6          & 7.8          & 7.7          \\
		0.1                & 5.0                & 7.8          & 7.9          & \textbf{7.4} & 7.7          & 7.8          & 7.7          \\ \midrule
		0.2                & 1.0                & 12.5         & 11.3         & 14           & 10.3         & 8.9          & 8.6          \\
		0.2                & 2.0                & 7.8          & \textbf{7.3} & 7.7          & 7.6          & \textbf{7.5} & 7.4          \\
		0.2                & 5.0                & \textbf{7.5} & 7.6          & 8.1          & \textbf{7.5} & 7.6          & 7.6          \\ \midrule
		0.5                & 1.0                & 36.8         & 26.9         & 40           & 28.1         & 14.4         & 10.6         \\
		0.5                & 2.0                & 10.9         & 9.9          & 12.2         & 9.5          & 8            & \textbf{7.3} \\
		0.5                & 5.0                & 8.1          & 7.8          & 8.4          & 7.8          & 7.6          & 7.5          \\ \bottomrule
	\end{tabular}
\end{table*}

\begin{table}[!t]	
	\caption{AISHELL-1: The evaluation of the LMs.}
	\centering
	\begin{tabular}{lccc}
		\toprule
		\multicolumn{1}{c}{\multirow{2}{*}{LM}} & \multicolumn{2}{c}{(pseudo) PPL/ACC}        & \multicolumn{1}{c}{\multirow{2}{*}{Model}}   \\ \cmidrule(l){2-3} 
		\multicolumn{1}{c}{}                    & Dev. Trans.         & Test Trans.           & \multicolumn{1}{c}{Size}                     \\ \midrule
		LSTM (Trans.)                           & 59.82 / 0.31         & 57.09 / 0.32         & 25.5M                                        \\
		LSTM (Ext.)                             & 35.12 / 0.36         & 33.84 / 0.37         & 25.5M                                        \\
		Transformer (Trans.)                    & 55.77 / 0.32         & 52.96 / 0.33         & 27.9M                                        \\
		Transformer (Ext.)                      & 28.50 / 0.39         & 27.66 / 0.40         & 27.9M                                        \\
		COR (Trans.)                            & 14.67 / 0.51         & 13.83 / 0.52         & 32.6M                                        \\
		COR (Ext.)                              & \textbf{5.59 / 0.63} & \textbf{5.48 / 0.65} & 32.6M                                        \\ \bottomrule
	\end{tabular}
	\label{tab:aishell1-acc}	
\end{table}

We employ several strategies for avoiding overfitting and achieving strong baselines. We use dropout with a rate of $0.1$. We use the SpecAugment \cite{park2019specaugment} strategy for data augmentation, but without time warping. It randomly masks out some inputted features with means. The frequency mask width is set to $27$, and the time mask width is set to $40$. Both frequency masking and time masking are processed twice.

We use the Adam algorithm \cite{kingma2014adam} as the optimizer. We follow the warm-up learning rate schedule \cite{vaswani2017attention}: 
\begin{equation}
\alpha = D^{-0.5}  \cdot \text{min} (step^{-0.5}, step \cdot warmup^{-1.5}), 
\end{equation}
where $D$ is the model dimensionality, i.e., 512, $step$ is the optimization step, and $warmup$ is the warm-up step. We set the warm-up step to $20000$. Each batch contains about $150$ seconds of speech data. $12$ iterations of gradients are accumulated for simulating big batch \cite{ott2018scaling}. For the different datasets, we train the model for different numbers of epochs, which are described in the experimental result part. All the AED models are trained from scratch. For decoding, we use beam search with width $5$.

\textbf{LMs}. We train three types of LMs: a long short-term memory (LSTM) based recurrent neural network LM, a transformer based LM \cite{irie2019language}, and the proposed COR. LSTM LM and transformer LM are autoregressive models, and COR leverages both the left context and the right context. All the LMs use the same vocabulary with the corresponding AED model.

\begin{table}[!t]	
	\caption{AISHELL-1: CERs on the test set.}
	\label{tab:aishell1-test}
	\centering
	\begin{threeparttable}
		\begin{tabular}{lcccc}
			\toprule
			\multicolumn{1}{c}{}                                            & $\lambda$ & $T$   & CER\%    &  \makecell[c]{Model\\Size} \\ \midrule
			KALDI (nnet3) * $\dagger$ $\ddagger$                            & -      & -        & 8.6      &           -                \\
			KALDI (chain) * $\dagger$ $\ddagger$                            & -      & -        & 7.4      &           -                \\
			LAS \cite{sun2019adversarial}                                   & -      & -        & 10.5     &           -                \\
			ESPnet (Transformer) $\dagger$ \cite{karita2019comparative}     & -      & -        & 6.7      &           -                \\
			ESPnet (Conformer) $\dagger$ \cite{guo2020recent}\cite{Gulati2020} \footnotemark & -      & -        & 4.7      &           -                \\
			Fan et al. \cite{fan2019unsupervised}                           & -      & -        & 6.7      &           -                \\
			An et al. $\dagger$ \cite{an2019cat}                            & -      & -        & 6.3      &           -                \\ \midrule
			Transformer (baseline)                                          & -      & -        & 7.6      &           67.5M            \\
			w/ Label Smoothing                                              & 0.1    & -        & 6.7      &           67.5M            \\
			w/ LSTM LM (Trans.)                                             & 0.2    & 5        & 6.5      &           67.5M            \\
			w/ LSTM LM (Ext.)                                               & 0.2    & 2        & 6.3      &           67.5M            \\
			w/ Transformer LM (Trans.)                                      & 0.1    & 5        & 6.5      &           67.5M            \\
			w/ Transformer LM (Ext.)                                        & 0.2    & 5        & 6.4      &           67.5M            \\
			w/ COR (Trans.)                                                 & 0.2    & 2        & 6.2      &           67.5M            \\
			w/ COR (Ext.)                                                   & 0.5    & 2        & \textbf{5.8} &       67.5M            \\ \bottomrule
		\end{tabular}
		\begin{tablenotes}	
			\footnotesize
			\item[*] from the KALDI official repository.
			\item[$\dagger$] with speed perturbation based data augmentation.
			\item[$\ddagger$] with i-vector based speaker adaptation.
		\end{tablenotes}
	\end{threeparttable}
\end{table}

\begin{table}[!t]	
	\caption{AISHELL-1: Comparisons with shallow fusion.}
	\centering
	\begin{tabular}{lcl}
		\toprule
		\multicolumn{1}{c}{Model} & CER\% & \makecell{Model\\Size}    \\ \midrule
		baseline                  & 7.6 & 67.5M         \\
		baseline + SF /w Tr.      & 6.4 & 67.5M + 27.9M \\
		LS                        & 6.7 & 67.5M         \\
		LS + SF /w Tr.            & 5.9 & 67.5M + 27.9M \\
		LST /w LSTM               & 6.3 & 67.5M         \\
		LST /w LSTM + SF w/ Tr.   & 7.2 & 67.5M + 27.9M \\
		LST /w Tr.                & 6.4 & 67.5M         \\
		LST /w Tr. + SF w/ Tr.    & \textbf{5.8} & 67.5M + 27.9M \\
		LST /w COR                & \textbf{5.8} & 67.5M         \\
		LST /w COR + SF w/ Tr.    & 6.2 & 67.5M + 27.9M \\ \bottomrule
	\end{tabular}
	\begin{tablenotes}
	  \footnotesize
	  \item LS is short for label smoothing. SF is short for shallow fusion. Tr. is short for transformer.
	\end{tablenotes}
	\label{tab:aishell1-sf}	
\end{table}

The LSTM LM has two layers. Each layer has $1024$ LSTM cells. We use the stochastic gradient descent (SGD) algorithm to train the LM for $20$ epochs. The batch size is set to $128$. The initial learning rate is $0.1$. After each epoch, the network is evaluated on the development set, and if the loss relative reduction is smaller than $10\%$, the learning rate is halved.

For the transformer LM, the model dimensionality is $512$. The number of heads is $8$, and the dimensionality of the position-wise feedforward module is $2048$. The number of the transformer layers is $5$. It also uses Adam algorithm and the transformer learning rate schedule. The warm-up step is $16000$. The models are trained for $20$ epochs. The batch size is set to $128$.

For COR model, the model dimensionality is also $512$, i.e., $D$ of ${U}$ in \autoref{eq:ori_att}. The number of transformer blocks for the forward stack and the backward stack is $5$. Each block has $8$ heads for attention. And the dimensionality of the position-wise feedforward module is also $2048$. The training schedule is the same as the transformer LM.

We evaluate each LM on the development set after each epoch during training, and the checkpoint which has the lowest loss is used as the final model.

\textbf{Evaluation}. For the LMs, because PPL cannot be computed for bidirectional language models, we compute ``pseudo'' PPL \cite{chen2017future} for the comparison purpose. To evaluate the effectiveness of the COR, we compute the cloze completion accuracy:
\begin{equation}
\text{ACC} = \frac{\#Corr}{\#Total}.
\end{equation} 
where $\#Total$ is the total number of tokens in the corpus, and $\#Corr$ is the total number tokens which the model predicts right. We use these two metrics to evaluate the LM performance.

And then we evaluate performance on the ASR task. We mainly focus on the improvements carried by the proposed methods. We use character error rate (CER) to evaluate the performance on the two Chinese datasets.

\subsection{Experimental Results on AISHELL-1}
\textbf{LM Evaluation}. We first evaluate the LM performance. \autoref{tab:aishell1-acc} shows the evaluation of the LMs. We can see that for each type of the LM, the external text significantly improves performance. Because variant sentences in the large-scale external text have more knowledge than the transcriptions. And more sentences make the LM avoid overfitting. The transformer based models outperform the LSTM models on both PPL and accuracy. Because the deep transformers directly model long dependency among the tokens, it shows more powerful ability for language modeling.

The LSTM LMs and the transformer LMs are unidirectional, i.e., they only use the left context to predict a token. So, the performance for predicting a token is limited. Different from these LMs, COR predict a token using both the left and the right context. Although only trained with the transcriptions, the accuracy is higher than the LSTM LMs and the transformer LMs. And the pseudo PPL is much reduced. With the external text, the performance of the COR is much increased. The COR trained on the external text achieves accuracy of $65\%$ on the text transcriptions, which is much better than the others.

\textbf{ASR Evaluation}. We then evaluate the LST training for ASR with different LMs. We first select the hyper parameters $\lambda$ and $T$ in \autoref{eq:loss} on the development set. For saving time, we train each model for $50$ epochs, and use the last checkpoint to evaluate the hyper parameters. We evaluate the typical values $0.1$, $0.2$, and $0.5$ for $\lambda$, and $1.0$, $2.0$, and $5.0$ for $T$. So, the number of combinations of the two hyper parameters is $9$. The results are shown in \autoref{tab:aishell1-hyper}. The hyper parameter $\lambda$ controls the ratio of $L_{LST}$, and $T$ controls the smoothness of the probability distributions provided by the teacher LM. 

\footnotetext{Conformer is a newly proposed module which introduces convolution into the self-attention module. }

From \autoref{tab:aishell1-hyper}, we can see that the proper hyperparameters depend on the teacher LM. A larger $\lambda$ introduces more knowledge from the LM. A larger $T$ makes the probability distributions of the LM smoother. For an LM with lower accuracy, it introduces more noise into the soft labels. Thus, a larger $T$ can be used to smooth the probability distribution. And $\lambda$ should not be too large so that the perturbations of the soft labels will not be too large.

Then, we use the selected hyperparameters corresponding to each LM to train the AED model for the whole $80$ epochs, and average the last $10$ models as the final model. The evaluation on the test set is shown in \autoref{tab:aishell1-test}. In our experiments, we do not use speed perturbation for data augmentation. As mentioned in \autoref{sec:rel}, label smoothing can be seen as zero-gram LM. We also report the result with label smoothing. The CER of the baseline model trained with cross-entropy is $7.6\%$. With LST training with different LMs, the performance are significantly improved. With the proposed COR LM which leverages both the left context and the right context, the performance is better than other unidirectional LM.  With COR trained on the external text, the CER is $5.8\%$. Compared with the baseline, the relative reduction of CER is $23.7\%$.

We analyze the impact of LST in two aspects. First, $L_{LST}$ introduces the uncertainty of the token labels. Specifically, when the LM is trained on the training transcriptions, the knowledge source of the LM is the same as the training transcriptions. LST training does not introduce external knowledge. But the soft labels introduce noise in expectation form to the loss as regularization. Second, when the LM is trained on the external text, the knowledge source of the LM is different from the training transcriptions. In this case, LST not only introduces uncertainty but also transfers knowledge from the external text with teacher-student learning. In addition, the COR models provide probability distributions given two-side context, and LST makes the AED model approach to the distributions. This introduces ``future'' knowledge from the ``future'' context in the token sequence.

To better understand the proposed method, we further use the shallow fusion technique to improve the performance of the models. Concretely, the scores of the LM are combined with an interpolation coefficient during beam-search. Thus, the criterion is as follows:
\begin{equation}
Y^* = \mathop{\arg\max}_{Y}(\log P_{AED}(Y|X) + \gamma\log P_{LM}(Y)).
\end{equation}
We use Transformer LM trained on the external text, which has the best PPL in our experiment. Note that COR is a bidirectional language model, so it cannot be used in shallow fusion. \autoref{tab:aishell1-sf} shows the results. LST is the proposed method. ``SF'' means Shallow Fusion. ``LSTM'' stands for the LSTM LM. ``Tr.'' stands for the transformer LM trained on the external text. COR is also trained on the external text. LS means label smoothing. The $\gamma$ is set to 0.1. We can see that with shallow fusion, the performance is significantly improved. However, because the whole system includes an LM, the size of the system becomes larger. We can also find some interesting phenomena in the table. First, the ``baseline + SF /w Tr.'' system has the same performance with ``LST /w Tr.'' system. It can be explained by that shallow fusion is an ensemble method during inference and the knowledge distillation is an implicit ensemble method during training \cite{hinton2015distilling,mun2018learning}. They show the same performance here. However, shallow fusion increases the size of system during inference. ``LST /w COR'' achieves the best performance. A strange phenomenon is that shallow fusion does not further improve the performance for ``LST /w COR'' and ``LST /w LSTM''. A possible reason is the discrepancy between the different LMs used in LST and shallow fusion.

\subsection{Experiments on AISHELL-2}
We extend the experiments on large-scale dataset AISHELL-2, which contains about $1000$ hours of speech data. The model structures are the same as previous AISHELL-1 experiment. The difference between the model sizes in AISHELL-2 experiments and AISHELL-1 experiments is caused by different vocabulary.

\textbf{LM Evaluation}. \autoref{tab:aishell2-acc} shows the (pseudo) PPL and accuracy of each LM. The COR models achieve significantly better accuracy. And with the external text, the COR achieves the highest accuracy.

\begin{table}[!t]	
	\caption{AISHELL-2: The evaluation of the LMs.}
	\centering
\begin{tabular}{lccc}
	\toprule
	\multicolumn{1}{c}{\multirow{2}{*}{LM}} & \multicolumn{2}{c}{(pseudo) PPL/ACC}   & \multicolumn{1}{c}{\multirow{2}{*}{Model}}                 \\ \cmidrule(l){2-3} 
	\multicolumn{1}{c}{}                    & Dev.Trans.        & Test Trans.        & \multicolumn{1}{c}{Size}                    \\ \midrule
	LSTM (Trans.)                           & 62.58 / 0.30      & 64.96 / 0.30       & 27.6M                   \\
	LSTM (Ext.)                             & 47.99 / 0.32      & 49.83 / 0.32       & 27.6M                  \\
	Transformer (Trans.)                    & 56.18 / 0.31      & 59.45 / 0.31       & 29.0M                  \\
	Transformer (Ext.)                      & 39.97 / 0.35      & 41.35 / 0.34       & 29.0M                   \\
	COR (Trans.)                            & 12.97 / 0.51      & 14.04 / 0.50       & 33.7M                  \\
	COR (Ext.)                              & \textbf{7.61 / 0.59}       & \textbf{7.98 / 0.58} & 33.7M     \\ \bottomrule
\end{tabular}
	\label{tab:aishell2-acc}	
\end{table}

\begin{table}[!t]	
	\caption{AISHELL-2: CERs on the test set.}
	\label{tab:aishell2-test}
	\centering
	\begin{threeparttable}
		\begin{tabular}{lccccc}
			\toprule
			                           & \multirow{2}{*}{$\lambda$} & \multirow{2}{*}{$T$} & \multicolumn{3}{c}{CER\%} \\ \cmidrule(l){4-6} 
			\multicolumn{1}{c}{}       &     &   & iPhone       & Android & HiFi         \\ \midrule
			KALDI (chain) $\dagger$ \cite{du2018aishell}  & -   & - & 8.8          & 9.6     & 10.9         \\
			LAS \cite{sun2019adversarial}                 & -   & - & 9.2          & 9.7     & 10.3         \\
			ESPnet (Transformer) *     & -   & - & 7.5          & 8.9     & 8.6          \\ 
			ESPnet (Conformer) \cite{guo2020recent}\cite{Gulati2020}    & -   & - & 6.8          & 7.6     & 7.4          \\\midrule
			Transformer (baseline)         & -   & - & 7.1          & 8.0     & 8.2          \\
			w/ Label Smooth            & 0.1 & - & 6.8          & 7.3     & 7.6          \\
			w/ LSTM LM (Trans.)        & 0.2 & 5 & 6.8          & 7.2     & 7.7          \\
			w/ LSTM LM (Ext.)          & 0.2 & 2 & 6.5          & 7.0     & 7.4          \\
			w/ Transformer LM (Trans.) & 0.1 & 5 & 6.6          & 7.2     & 7.8          \\
			w/ Transformer LM (Ext.)   & 0.2 & 5 & 6.4          & 7.1     & 7.5          \\
			w/ COR (Trans.)            & 0.2 & 2 & 6.0          & \textbf{6.8}     & 7.3          \\
			w/ COR (Ext.)              & 0.5 & 2 & \textbf{5.7} & \textbf{6.8}     & \textbf{7.1} \\ \bottomrule
		\end{tabular}
		\begin{tablenotes}	
			\footnotesize
			\item[*] from the ESPnet official repository.
			\item[$\dagger$] with i-vector based speaker adaptation. 
			
			The numbers of parameters of the our implemented transformers in this group of experiments are 68.5M.
		\end{tablenotes}
	\end{threeparttable}
\end{table}

\begin{table}[!t]	
	\caption{AISHELL-2: Comparisons with shallow fusion.}
	\centering
\begin{tabular}{lcccl}
	\toprule
	\multicolumn{1}{c}{\multirow{2}{*}{Model}}     & \multicolumn{3}{c}{CER\%}      &  \multicolumn{1}{c}{\multirow{2}{*}{\makecell{Model\\Size}}}                \\
		\multicolumn{1}{c}{}                       & iPh.  & An. & HiFi       &                                                               \\ \midrule
		baseline                                   & 7.1     & 8.0     & 8.2        &  68.5M                                                     \\
		baseline + SF /w Tr.                       & 6.6     & 7.3     & 7.6        &  68.5M + 29.0M                                             \\
		LS                                         & 6.8     & 7.3     & 7.6        &  68.5M                                                             \\
		LS + SF /w Tr.                             & 5.9     & \textbf{6.7}     & \textbf{7.1} & 68.5M + 29.0M  \\
		LST /w LSTM                                & 6.5     & 7.0     & 7.4                   & 68.5M                 \\
		LST /w LSTM + SF w/ Tr.                    & 7.4     & 7.8     & 8.2                   & 68.5M + 29.0M    \\
		LST /w Tr.                                 & 6.4     & 7.1     & 7.5                   & 68.5M                   \\
		LST /w Tr. + SF w/ Tr.                     & 6.5     & 7.0     & 7.3                   & 68.5M + 29.0M                         \\
		LST /w COR                                 & \textbf{5.7}     & 6.8     & \textbf{7.1} & 68.5M  \\
		LST /w COR + SF w/ Tr.                     & 5.9     & 7.0     & 7.3                   & 68.5M + 29.0M      \\ \bottomrule
	\end{tabular}
	\begin{tablenotes}	
	\footnotesize
	\item IPh. is short for iPhone. An. is short for Android. LS is short for label smoothing. SF is short for shallow fusion. Tr. is short for transformer.
	\end{tablenotes}
	\label{tab:aishell2-sf}	
\end{table}

\begin{figure*}[!t]\centering
	\subfloat[Cross-Entropy] 
	{ \label{fig:nosm}
		\includegraphics[width=0.45\columnwidth]{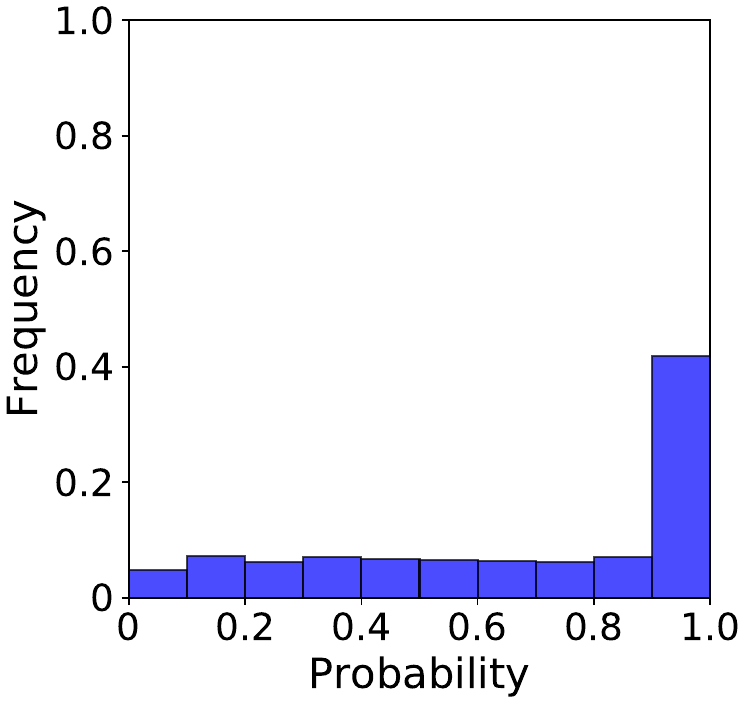}
	}		
	\quad
	\subfloat[Label Smoothing (LST with the zero-gram LM)]  
	{ \label{fig:ls01}
		\includegraphics[width=0.45\columnwidth]{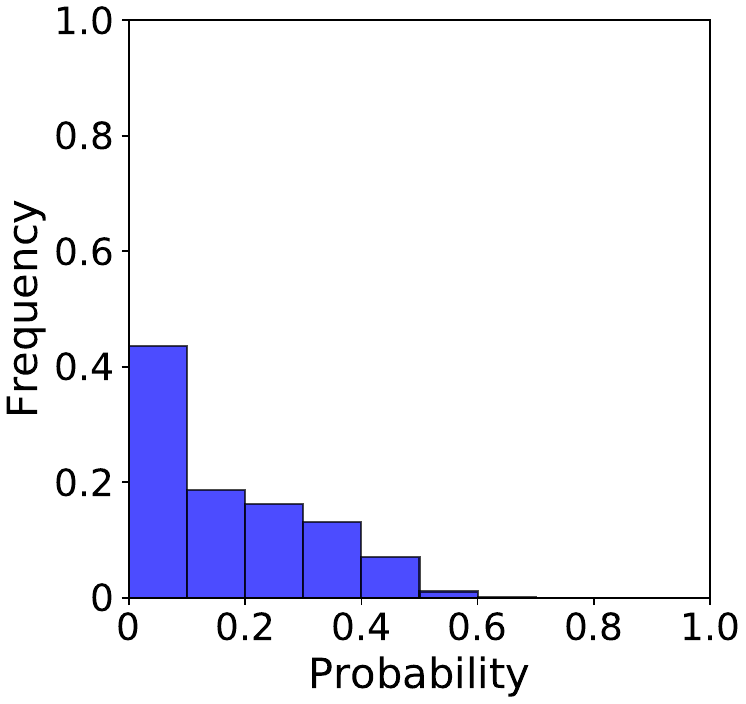}
	}
	\quad
	\subfloat[LST /w LSTM LM (Trans.)]  
	{ \label{fig:lstmtrans}
		\includegraphics[width=0.45\columnwidth]{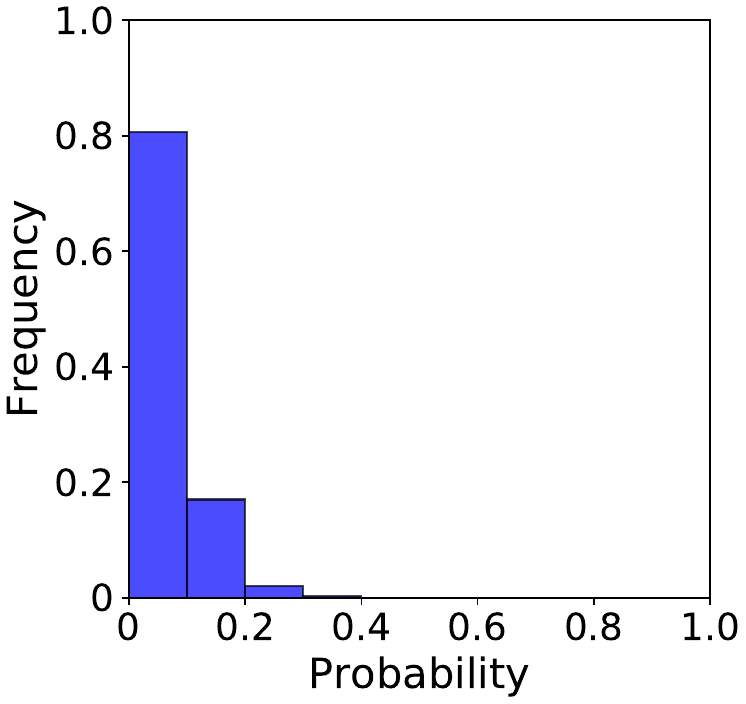}
	}
	\quad
	\subfloat[LST /w LSTM LM (Ext.)]  
	{ \label{fig:lstmext}
		\includegraphics[width=0.45\columnwidth]{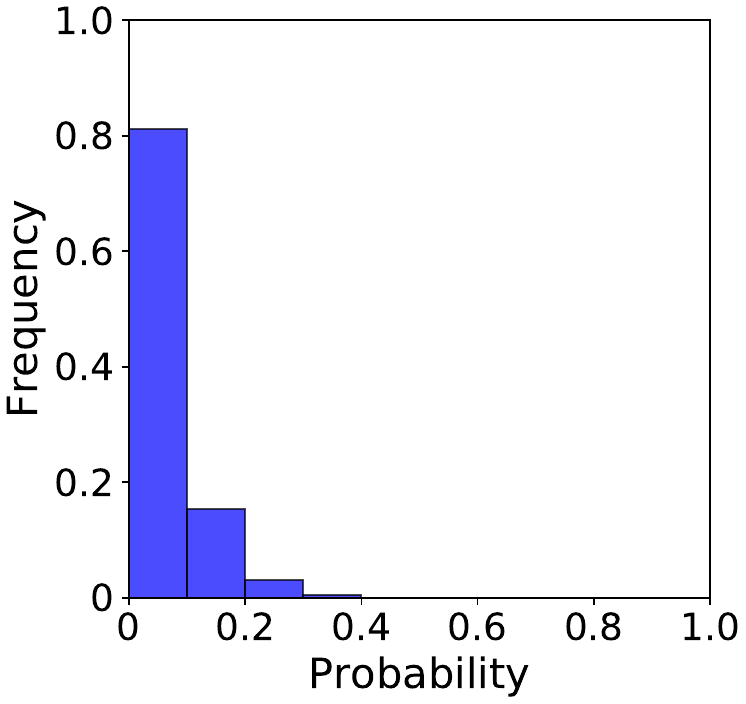}
	}
	
	\subfloat[LST /w Transformer LM (Trans.)]
	{ \label{fig:trtrans}
		\includegraphics[width=0.45\columnwidth]{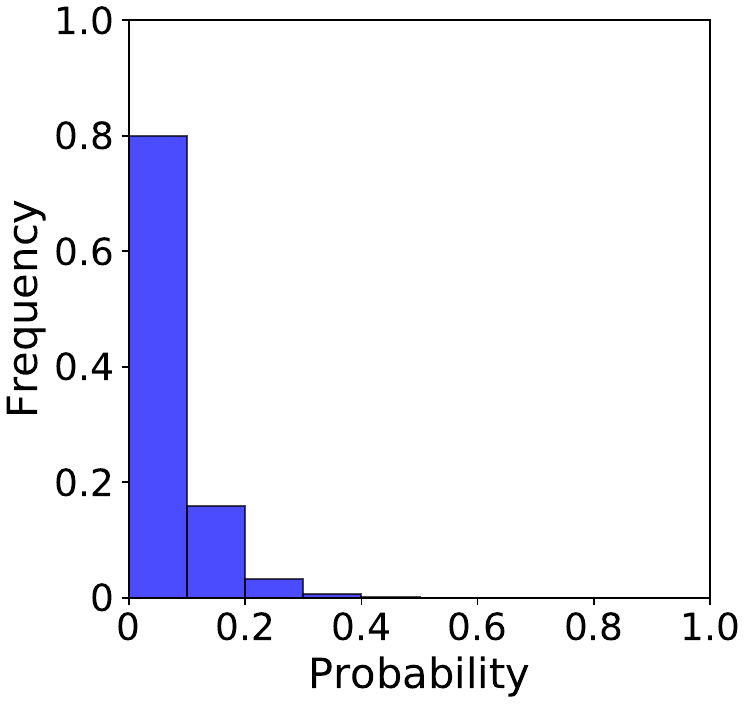}
	}		
	\quad
	\subfloat[LST /w Transformer LM (Ext.)]  
	{ \label{fig:trext}
		\includegraphics[width=0.45\columnwidth]{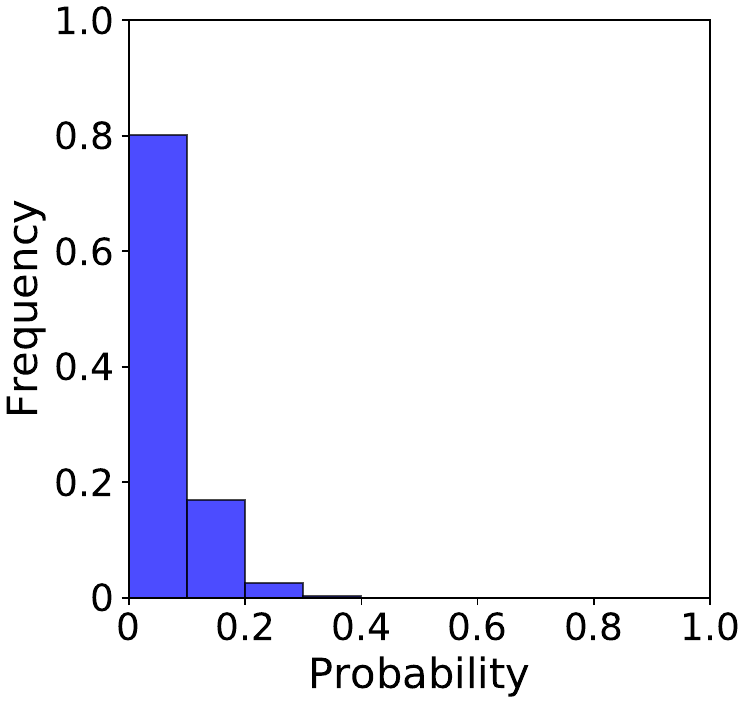}
	}
	\quad
	\subfloat[LST /w COR (Trans.)]  
	{ \label{fig:cortrans}
		\includegraphics[width=0.45\columnwidth]{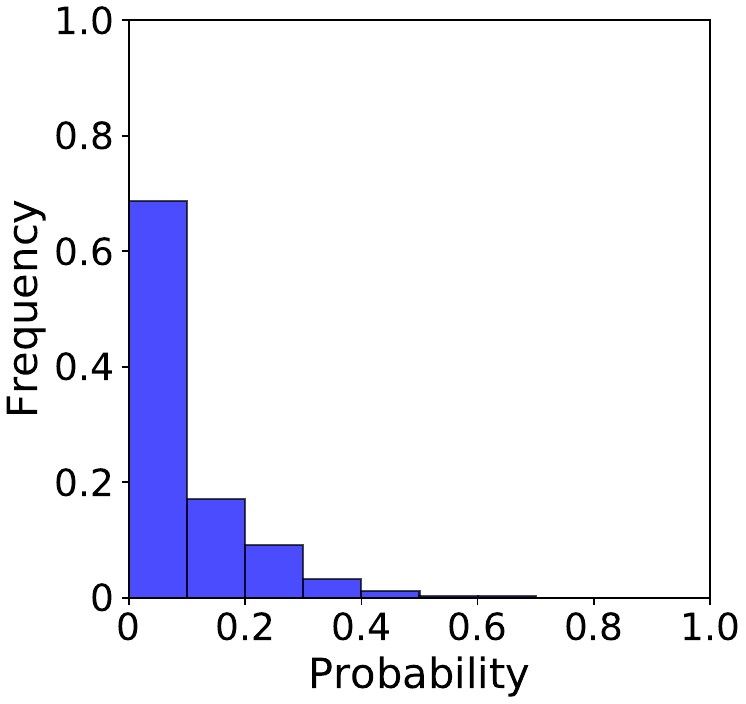}
	}
	\quad
	\subfloat[LST /w COR (Ext.)]  
	{ \label{fig:corext}
		\includegraphics[width=0.45\columnwidth]{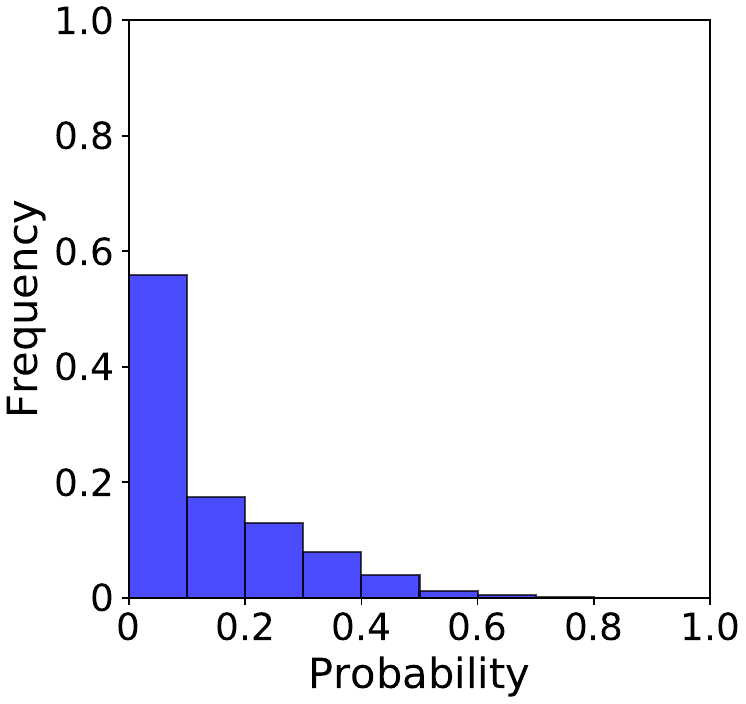}
	}

	\subfloat[Cross-Entropy + SF /w Transformer LM (Ext.)]
	{ \label{fig:nosm_sf}
		\includegraphics[width=0.34\columnwidth]{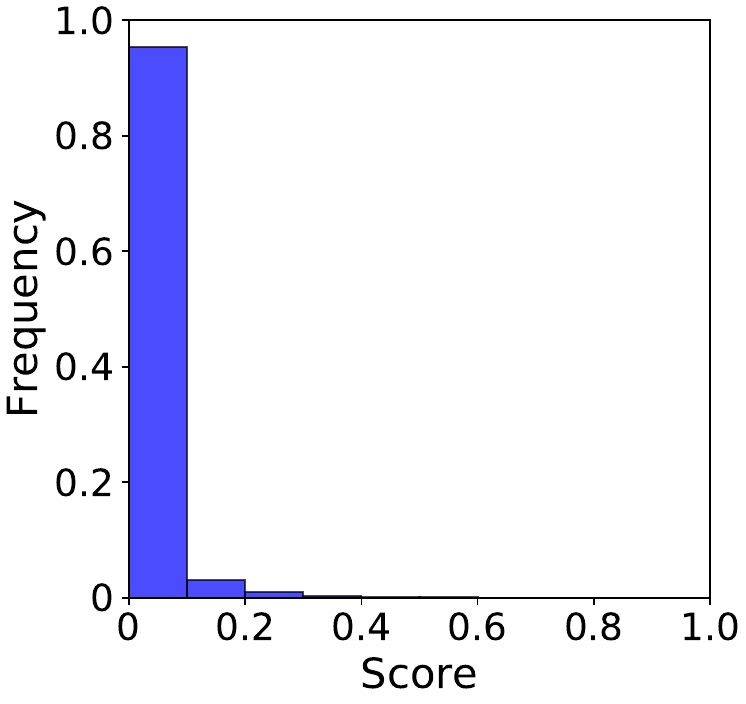}
	}		
	\quad
	\subfloat[Label Smoothing + SF /w Transformer LM (Ext.)]  
	{ \label{fig:ls01_sf}
		\includegraphics[width=0.34\columnwidth]{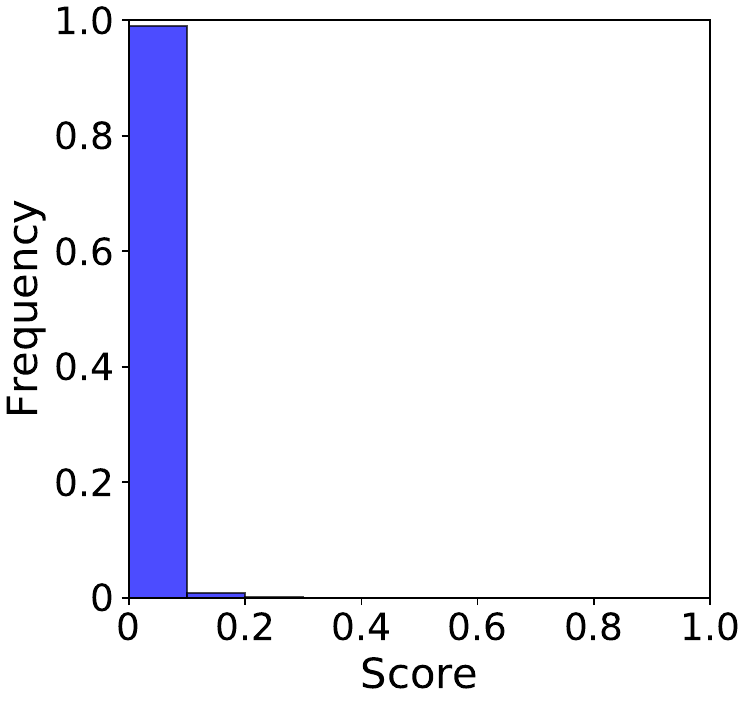}
	}
	\quad
	\subfloat[LST /w LSTM LM (Ext.) + SF /w Transformer LM (Ext.)]  
	{ \label{fig:lstmext_sf}
		\includegraphics[width=0.34\columnwidth]{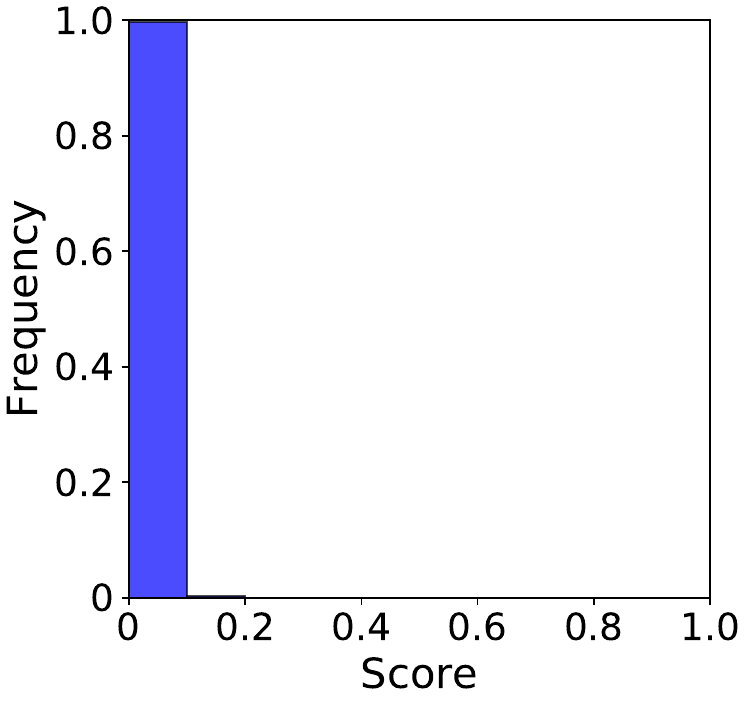}
	}
	\quad
	\subfloat[LST /w Transformer LM (Ext.) + SF /w Transformer LM (Ext.)]  
	{ \label{fig:trext_sf}
		\includegraphics[width=0.34\columnwidth]{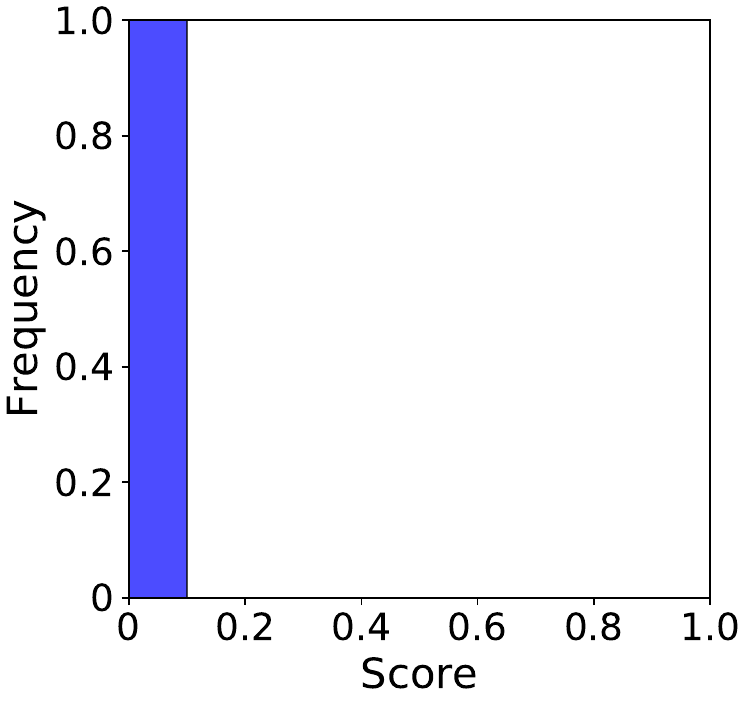}
	}
	\quad
	\subfloat[LST /w COR (Ext.)  + SF /w Transformer LM (Ext.)]  
	{ \label{fig:corext_sf}
		\includegraphics[width=0.34\columnwidth]{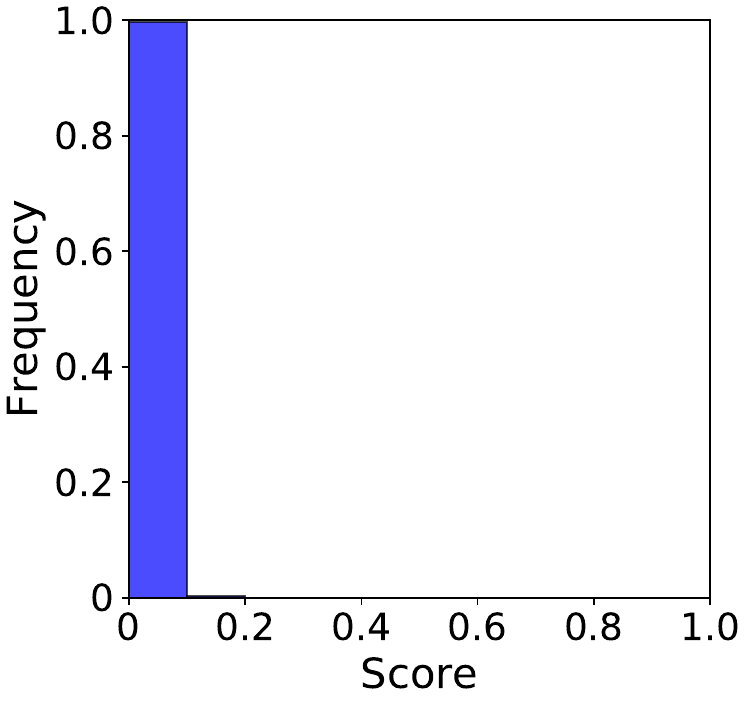}
	}

	\caption{Histograms the probabilities of the top-1 hypotheses AISHELL-1 test set. The hypotheses are generated by the models trained with different loss functions or shallow fusion. The models are the same ones in \autoref{tab:aishell1-test} and \autoref{tab:aishell1-sf}.} 
	\label{fig:hisograms}
\end{figure*}

\textbf{ASR Evaluation}. Because training models on the 1000-hour AISHELL-2 dataset costs much time, we directly use the best hyperparameters selected on AISHELL-1. We train the models for $40$ epochs and average the parameters of the models at the last $10$ epochs. The experimental results are shown in \autoref{tab:aishell2-test}. Similar to the results on AISHELL-1, with LST training, the performance of the transformer models are improved. And when the LM is COR which is trained on the external text, the model achieves the lowest CER on the test set. Compared with the baseline model, which is trained with cross-entropy, the best model which is trained with LST achieves $13\%$ to $19\%$ relative reduction. Similarly, we also compare the proposed method with shallow fusion, which is shown in \autoref{tab:aishell2-sf}. The similar phenomena with AISHELL-1 are observed. We can see that the proposed method does not increase model size during inference, compared with shallow fusion. We also find that shallow fusion does not further improve the performance for ``LST /w COR'' and ``LST /w LSTM'' in this group of experiments. But it can improve the model with label smoothing and the model trained with transformer LM. This is the same with previous AISHELL-1 experiments. We analyze that a possible reason is the discrepancy between the different LMs used in LST and shallow fusion.  

\vspace{-10pt}
\subsection{An Analysis of Probabilities of Hypotheses}
We observe that the models trained with different loss functions behave differently during decoding. Specifically, the log probabilities of the top-1 hypotheses which are generated with different models are very different in value. To better understand the behaviors of the AED models trained with different losses, we draw the probability histograms of top-1 hypotheses on AISHELL-1 test set. The models are trained with the best hyperparameters which are selected in the AISHELL-1 experiments.

From \autoref{fig:hisograms}, we can see that the histogram of the model trained with only cross-entropy (\autoref{fig:nosm}) is very different from other models which are trained with LST. Specifically, the most probabilities of the top-1 hypotheses are in $[0.9, 1.0]$ in \autoref{fig:nosm}. But in the other histograms, the most probabilities of the top-1 hypotheses are in $[0, 0.1]$. The histograms of the models which use the LMs with the same structures as teachers are similar. The histograms of the models trained with COR are relatively smoother (\autoref{fig:cortrans} and \autoref{fig:corext}). And these two models achieve better performance.

The model trained only with cross-entropy decodes the hypotheses with very high confidence scores. This overconfident phenomenon is a manifestation of overfitting. The opportunity of selecting the correct hypothesis which is not top-1 will be smaller during decoding. LST training alleviates this problem. With LST, there are few hypotheses with probabilities in $[0.9, 1.0]$.

The histograms partially show the probability spaces estimated by the AED models. Computing distribution of the probabilities of all sequences is intractable, so we use the probability of the top-1 to partially show the space. From \autoref{fig:nosm}, we can imagine that given a speech sequence, the probability distribution over the token sequences, which are estimated by the AED model, concentrates on some examples. LST training smooths the probability space estimated by an AED model.

In the last row of \autoref{fig:hisograms}, we show the histograms of scores of the shallow fusion results. Note that the score of a shallow fusion result is obtained with the following equation:
\begin{equation}
\begin{split}
\text{Score} &= \exp(\log P_{AED}(Y|X) + \gamma\log P_{LM}(Y)) \\
			 &= P_{AED}(Y|X)P_{LM}(Y)^\gamma.
\end{split}
\end{equation}
Thus, this score is not a probability and cannot be compared with probabilities in the above figures directly. But we still can find that shallow fusion smooths the score distributions of the hypotheses.

Shallow fusion is an ensemble method during inference stage. LST is an ensemble method during training stage \cite{hinton2015distilling,mun2018learning}. \autoref{fig:hisograms} shows that these two methods similar features: both of them smooth the score spaces of the hypotheses. And both of them improve the performance of an AED model. These results provide intuitive materials for analysis of ensemble methods of sequence models. 

\section{Conclusions and Future Work}
\label{sec:conc}
This paper proposes a method LST to integrate knowledge from external text-only data to an attention-based encoder-decoder model. First, the knowledge in the external text-only data is represented in an LM. Then, the teacher-student learning is used to transfer the knowledge from the LM to the AED model based ASR system. In order to leverage the whole context rather than only the left context in a sentence, we propose a new LM called COR, which is based on self-attention mechanism. The experiments on two public Chinese datasets demonstrate the effectiveness of LST. In the future, we will explore how to set the hyperparameters automatically. And we will explore more effective AED model, such as conformer\cite{Gulati2020}, to improve the performance. And the theoretical analysis of knowledge distillation of sequence problems is also worthy to investigate.

\section{Acknowledgment}
The authors are grateful to the anonymous reviewers for their invaluable comments that improve the completeness and readability of this paper.
\ifCLASSOPTIONcaptionsoff
  \newpage
\fi



%


\begin{thebibliography}{10}
\providecommand{\url}[1]{#1}
\csname url@samestyle\endcsname
\providecommand{\newblock}{\relax}
\providecommand{\bibinfo}[2]{#2}
\providecommand{\BIBentrySTDinterwordspacing}{\spaceskip=0pt\relax}
\providecommand{\BIBentryALTinterwordstretchfactor}{4}
\providecommand{\BIBentryALTinterwordspacing}{\spaceskip=\fontdimen2\font plus
\BIBentryALTinterwordstretchfactor\fontdimen3\font minus
  \fontdimen4\font\relax}
\providecommand{\BIBforeignlanguage}[2]{{%
\expandafter\ifx\csname l@#1\endcsname\relax
\typeout{** WARNING: IEEEtran.bst: No hyphenation pattern has been}%
\typeout{** loaded for the language `#1'. Using the pattern for}%
\typeout{** the default language instead.}%
\else
\language=\csname l@#1\endcsname
\fi
#2}}
\providecommand{\BIBdecl}{\relax}
\BIBdecl

\bibitem{yu2016automatic}
D.~Yu and L.~Deng, \emph{AUTOMATIC SPEECH RECOGNITION.}\hskip 1em plus 0.5em
  minus 0.4em\relax Springer, 2016.

\bibitem{vesely2013sequence}
K.~Vesel{\`y}, A.~Ghoshal, L.~Burget, and D.~Povey, ``Sequence-discriminative
  training of deep neural networks.'' in \emph{Interspeech}, vol. 2013, 2013,
  pp. 2345--2349.

\bibitem{sainath2013deep}
T.~N. Sainath, A.-r. Mohamed, B.~Kingsbury, and B.~Ramabhadran, ``Deep
  convolutional neural networks for lvcsr,'' in \emph{2013 IEEE international
  conference on acoustics, speech and signal processing}.\hskip 1em plus 0.5em
  minus 0.4em\relax IEEE, 2013, pp. 8614--8618.

\bibitem{zeyer2017comprehensive}
A.~Zeyer, P.~Doetsch, P.~Voigtlaender, R.~Schl{\"u}ter, and H.~Ney, ``A
  comprehensive study of deep bidirectional lstm rnns for acoustic modeling in
  speech recognition,'' in \emph{2017 IEEE International Conference on
  Acoustics, Speech and Signal Processing (ICASSP)}.\hskip 1em plus 0.5em minus
  0.4em\relax IEEE, 2017, pp. 2462--2466.

\bibitem{peddinti2015time}
V.~Peddinti, D.~Povey, and S.~Khudanpur, ``A time delay neural network
  architecture for efficient modeling of long temporal contexts,'' in
  \emph{Sixteenth Annual Conference of the International Speech Communication
  Association}, 2015.

\bibitem{chorowski2015attention}
J.~K. Chorowski, D.~Bahdanau, D.~Serdyuk, K.~Cho, and Y.~Bengio,
  ``Attention-based models for speech recognition,'' in \emph{Advances in
  neural information processing systems}, 2015, pp. 577--585.

\bibitem{bahdanau2016endtoend}
D.~Bahdanau, J.~Chorowski, D.~Serdyuk, P.~Brakel, and Y.~Bengio, ``End-to-end
  attention-based large vocabulary speech recognition,'' \emph{international
  conference on acoustics, speech, and signal processing}, pp. 4945--4949,
  2016.

\bibitem{chan2016listen}
W.~Chan, N.~Jaitly, Q.~Le, and O.~Vinyals, ``Listen, attend and spell: A neural
  network for large vocabulary conversational speech recognition,'' in
  \emph{2016 IEEE International Conference on Acoustics, Speech and Signal
  Processing (ICASSP)}.\hskip 1em plus 0.5em minus 0.4em\relax IEEE, 2016, pp.
  4960--4964.

\bibitem{kim2017joint}
S.~Kim, T.~Hori, and S.~Watanabe, ``Joint ctc-attention based end-to-end speech
  recognition using multi-task learning,'' in \emph{2017 IEEE international
  conference on acoustics, speech and signal processing (ICASSP)}.\hskip 1em
  plus 0.5em minus 0.4em\relax IEEE, 2017, pp. 4835--4839.

\bibitem{dong2018speech}
L.~Dong, S.~Xu, and B.~Xu, ``Speech-transformer: a no-recurrence
  sequence-to-sequence model for speech recognition,'' in \emph{2018 IEEE
  International Conference on Acoustics, Speech and Signal Processing
  (ICASSP)}.\hskip 1em plus 0.5em minus 0.4em\relax IEEE, 2018, pp. 5884--5888.

\bibitem{chiu2018state}
C.-C. Chiu, T.~N. Sainath, Y.~Wu, R.~Prabhavalkar, P.~Nguyen, Z.~Chen,
  A.~Kannan, R.~J. Weiss, K.~Rao, E.~Gonina \emph{et~al.}, ``State-of-the-art
  speech recognition with sequence-to-sequence models,'' in \emph{2018 IEEE
  International Conference on Acoustics, Speech and Signal Processing
  (ICASSP)}.\hskip 1em plus 0.5em minus 0.4em\relax IEEE, 2018, pp. 4774--4778.

\bibitem{graves2012sequence}
A.~Graves, ``Sequence transduction with recurrent neural networks,''
  \emph{arXiv preprint arXiv:1211.3711}, 2012.

\bibitem{rao2017exploring}
K.~Rao, H.~Sak, and R.~Prabhavalkar, ``Exploring architectures, data and units
  for streaming end-to-end speech recognition with rnn-transducer,'' in
  \emph{2017 IEEE Automatic Speech Recognition and Understanding Workshop
  (ASRU)}.\hskip 1em plus 0.5em minus 0.4em\relax IEEE, 2017, pp. 193--199.

\bibitem{peters2018deep}
M.~Peters, M.~Neumann, M.~Iyyer, M.~Gardner, C.~Clark, K.~Lee, and
  L.~Zettlemoyer, ``Deep contextualized word representations,''
  \emph{Proceedings of the 2018 Conference of the North {A}merican Chapter of
  the Association for Computational Linguistics: Human Language Technologies,
  Volume 1 (Long Papers)}, pp. 2227--2237, Jun. 2018.

\bibitem{devlin2018bert}
J.~Devlin, M.-W. Chang, K.~Lee, and K.~Toutanova, ``{BERT}: Pre-training of
  deep bidirectional transformers for language understanding,'' in
  \emph{Proceedings of the 2019 Conference of the North {A}merican Chapter of
  the Association for Computational Linguistics: Human Language Technologies,
  Volume 1 (Long and Short Papers)}.\hskip 1em plus 0.5em minus 0.4em\relax
  Minneapolis, Minnesota: Association for Computational Linguistics, Jun. 2019,
  pp. 4171--4186.

\bibitem{gulcehre2015on}
C.~Gulcehre, O.~Firat, K.~Xu, K.~Cho, L.~Barrault, H.~Lin, F.~Bougares,
  H.~Schwenk, and Y.~Bengio, ``On using monolingual corpora in neural machine
  translation,'' \emph{arXiv preprint arXiv:1503.03535}, 2015.

\bibitem{sriram2018cold}
A.~Sriram, H.~Jun, S.~Satheesh, and A.~Coates, ``Cold fusion: Training seq2seq
  models together with language models,'' pp. 387--391, 2018.

\bibitem{kannan2018an}
A.~{Kannan}, Y.~{Wu}, P.~{Nguyen}, T.~N. {Sainath}, Z.~{Chen}, and
  R.~{Prabhavalkar}, ``An analysis of incorporating an external language model
  into a sequence-to-sequence model,'' pp. 1--5828, 2018.

\bibitem{shan2019component}
C.~Shan, C.~Weng, G.~Wang, D.~Su, M.~Luo, D.~Yu, and L.~Xie, ``Component
  fusion: Learning replaceable language model component for end-to-end speech
  recognition system,'' in \emph{2019 IEEE International Conference on
  Acoustics, Speech and Signal Processing (ICASSP)}.\hskip 1em plus 0.5em minus
  0.4em\relax IEEE, 2019.

\bibitem{bengio2015scheduled}
S.~Bengio, O.~Vinyals, N.~Jaitly, and N.~Shazeer, ``Scheduled sampling for
  sequence prediction with recurrent neural networks,'' \emph{Advances in
  Neural Information Processing Systems 28}, pp. 1171--1179, 2015.

\bibitem{mimura2018forward}
M.~Mimura, S.~Sakai, and T.~Kawahara, ``Forward-backward attention decoder.''
  \emph{Proc. Interspeech 2018}, pp. 2232--2236, 2018.

\bibitem{zhou2019synchronous}
L.~Zhou, J.~Zhang, and C.~Zong, ``Synchronous bidirectional neural machine
  translation,'' \emph{Transactions of the Association for Computational
  Linguistics}, vol.~7, pp. 91--105, 2019.

\bibitem{liu-etal-2016-agreement-target}
L.~Liu, M.~Utiyama, A.~Finch, and E.~Sumita, ``Agreement on
  target-bidirectional neural machine translation,'' in \emph{Proceedings of
  the 2016 Conference of the North {A}merican Chapter of the Association for
  Computational Linguistics: Human Language Technologies}.\hskip 1em plus 0.5em
  minus 0.4em\relax Association for Computational Linguistics, Jun. 2016, pp.
  411--416.

\bibitem{zhang2019regularizing}
Z.~Zhang, S.~Wu, S.~Liu, M.~Li, M.~Zhou, and T.~Xu, ``Regularizing neural
  machine translation by target-bidirectional agreement,'' in \emph{Proceedings
  of the AAAI Conference on Artificial Intelligence}, vol.~33, 2019, pp.
  443--450.

\bibitem{zheng2019forward}
Y.~Zheng, J.~Tao, Z.~Wen, and J.~Yi, ``Forward–backward decoding sequence for
  regularizing end-to-end tts,'' \emph{IEEE Transactions on Audio, Speech, and
  Language Processing}, vol.~27, no.~12, pp. 2067--2079, 2019.

\bibitem{2006model}
C.~Buciluǎ, R.~Caruana, and A.~Niculescu-Mizil, ``Model compression,'' in
  \emph{Proceedings of the 12th ACM SIGKDD international conference on
  Knowledge discovery and data mining}, 2006, pp. 535--541.

\bibitem{hinton2015distilling}
G.~Hinton, O.~Vinyals, and J.~Dean, ``Distilling the knowledge in a neural
  network,'' \emph{arXiv preprint arXiv:1503.02531}, 2015.

\bibitem{romero2014fitnets}
A.~Romero, N.~Ballas, S.~E. Kahou, A.~Chassang, C.~Gatta, and Y.~Bengio,
  ``Fitnets: Hints for thin deep nets,'' \emph{arXiv preprint arXiv:1412.6550},
  2014.

\bibitem{li2014learning}
J.~Li, R.~Zhao, J.-T. Huang, and Y.~Gong, ``Learning small-size dnn with
  output-distribution-based criteria,'' in \emph{Fifteenth annual conference of
  the international speech communication association}, 2014.

\bibitem{li2017large}
J.~Li, M.~L. Seltzer, X.~Wang, R.~Zhao, and Y.~Gong, ``Large-scale domain
  adaptation via teacher-student learning,'' \emph{Proc. Interspeech 2017}, pp.
  2386--2390, 2017.

\bibitem{bai2019learn}
Y.~Bai, J.~Yi, J.~Tao, Z.~Tian, and Z.~Wen, ``Learn spelling from teachers:
  Transferring knowledge from language models to sequence-to-sequence speech
  recognition,'' \emph{Proc. Interspeech 2019}, pp. 3795--3799, 2019.

\bibitem{vaswani2017attention}
A.~Vaswani, N.~Shazeer, N.~Parmar, J.~Uszkoreit, L.~Jones, A.~N. Gomez,
  {\L}.~Kaiser, and I.~Polosukhin, ``Attention is all you need,'' in
  \emph{Advances in Neural Information Processing Systems}, 2017, pp.
  5998--6008.

\bibitem{taylor1953cloze}
W.~L. Taylor, ``“cloze procedure”: A new tool for measuring readability,''
  \emph{Journalism Bulletin}, vol.~30, no.~4, pp. 415--433, 1953.

\bibitem{mousa2017contextual}
A.~Mousa and B.~Schuller, ``Contextual bidirectional long short-term memory
  recurrent neural network language models: A generative approach to sentiment
  analysis,'' in \emph{Proceedings of the 15th Conference of the European
  Chapter of the Association for Computational Linguistics}, 2017, pp.
  1023--1032.

\bibitem{baevski2019cloze}
A.~Baevski, S.~Edunov, Y.~Liu, L.~Zettlemoyer, and M.~Auli, ``Cloze-driven
  pretraining of self-attention networks,'' \emph{Proceedings of the 2019
  Conference on Empirical Methods in Natural Language Processing and the 9th
  International Joint Conference on Natural Language Processing
  (EMNLP-IJCNLP)}, pp. 5360--5369, Nov. 2019.

\bibitem{dauphin2017language}
Y.~N. Dauphin, A.~Fan, M.~Auli, and D.~Grangier, ``Language modeling with gated
  convolutional networks,'' in \emph{Proceedings of the 34th International
  Conference on Machine Learning-Volume 70}.\hskip 1em plus 0.5em minus
  0.4em\relax JMLR. org, 2017, pp. 933--941.

\bibitem{he2016}
K.~He, X.~Zhang, S.~Ren, and J.~Sun, ``Deep residual learning for image
  recognition,'' in \emph{2016 {IEEE} Conference on Computer Vision and Pattern
  Recognition, {CVPR} 2016, Las Vegas, NV, USA, June 27-30, 2016}.\hskip 1em
  plus 0.5em minus 0.4em\relax {IEEE} Computer Society, 2016, pp. 770--778.

\bibitem{ba2016}
L.~J. Ba, J.~R. Kiros, and G.~E. Hinton, ``Layer normalization,'' \emph{arXiv
  preprint arXiv:1607.06450}, 2016.

\bibitem{kim2016sequence}
Y.~Kim and A.~M. Rush, ``Sequence-level knowledge distillation,''
  \emph{Proceedings of the 2016 Conference on Empirical Methods in Natural
  Language Processing}, pp. 1317--1327, Nov. 2016.

\bibitem{andres2018efficient}
J.~Andr{\'e}s-Ferrer, N.~Bodenstab, and P.~Vozila, ``Efficient language model
  adaptation with noise contrastive estimation and kullback-leibler
  regularization,'' \emph{Proc. Interspeech 2018}, pp. 3368--3372, 2018.

\bibitem{szegedy2016rethinking}
C.~Szegedy, V.~Vanhoucke, S.~Ioffe, J.~Shlens, and Z.~Wojna, ``Rethinking the
  inception architecture for computer vision,'' in \emph{Proceedings of the
  IEEE conference on computer vision and pattern recognition}, 2016, pp.
  2818--2826.

\bibitem{pereyra2017regularizing}
G.~Pereyra, G.~Tucker, J.~Chorowski, {\L}.~Kaiser, and G.~Hinton,
  ``Regularizing neural networks by penalizing confident output
  distributions,'' \emph{International Conference on Learning Representations
  Workshop}, 2017.

\bibitem{chorowski2017towards}
J.~Chorowski and N.~Jaitly, ``Towards better decoding and language model
  integration in sequence to sequence models,'' \emph{Proc. Interspeech 2017},
  pp. 523--527, 2017.

\bibitem{rosenfeld2001whole}
R.~Rosenfeld, S.~F. Chen, and X.~Zhu, ``Whole-sentence exponential language
  models: a vehicle for linguistic-statistical integration,'' \emph{Computer
  Speech \& Language}, vol.~15, no.~1, pp. 55--73, 2001.

\bibitem{chen2009shrinking}
S.~F. Chen, ``Shrinking exponential language models,'' in \emph{Proceedings of
  Human Language Technologies: The 2009 Annual Conference of the North American
  Chapter of the Association for Computational Linguistics}, 2009, pp.
  468--476.

\bibitem{amaya2001improvement}
F.~A. Amaya and J.-M. Bened{\'\i}, ``Improvement of a whole sentence maximum
  entropy language model using grammatical features,'' in \emph{Proceedings of
  the 39th Annual Meeting of the Association for Computational Linguistics},
  2001, pp. 10--17.

\bibitem{wang2017learning}
B.~Wang, Z.~Ou, and Z.~Tan, ``Learning trans-dimensional random fields with
  applications to language modeling,'' \emph{IEEE transactions on pattern
  analysis and machine intelligence}, vol.~40, no.~4, pp. 876--890, 2017.

\bibitem{arisoy2015bi}
E.~{Arisoy}, A.~{Sethy}, B.~{Ramabhadran}, and S.~{Chen}, ``Bidirectional
  recurrent neural network language models for automatic speech recognition,''
  in \emph{2015 IEEE International Conference on Acoustics, Speech and Signal
  Processing (ICASSP)}, 2015, pp. 5421--5425.

\bibitem{he2016on}
T.~{He}, Y.~{Zhang}, J.~{Droppo}, and K.~{Yu}, ``On training bi-directional
  neural network language model with noise contrastive estimation,'' in
  \emph{2016 10th International Symposium on Chinese Spoken Language Processing
  (ISCSLP)}, 2016, pp. 1--5.

\bibitem{chen2017investigating}
X.~Chen, A.~Ragni, X.~Liu, and M.~J. Gales, ``Investigating bidirectional
  recurrent neural network language models for speech recognition,'' in
  \emph{Proceedings of Interspeech 2017}.\hskip 1em plus 0.5em minus
  0.4em\relax International Speech Communication Association (ISCA), 2017, pp.
  269--273.

\bibitem{zhang2019ernie}
Z.~Zhang, X.~Han, Z.~Liu, X.~Jiang, M.~Sun, and Q.~Liu, ``Ernie: Enhanced
  language representation with informative entities,'' in \emph{Proceedings of
  the 57th Annual Meeting of the Association for Computational Linguistics},
  2019, pp. 1441--1451.

\bibitem{Yang2019XLNet}
Z.~Yang, Z.~Dai, Y.~Yang, J.~Carbonell, R.~R. Salakhutdinov, and Q.~V. Le,
  ``Xlnet: Generalized autoregressive pretraining for language understanding,''
  pp. 5753--5763, 2019.

\bibitem{bu2017aishell}
H.~Bu, J.~Du, X.~Na, B.~Wu, and H.~Zheng, ``Aishell-1: An open-source mandarin
  speech corpus and a speech recognition baseline,'' in \emph{2017 20th
  Conference of the Oriental Chapter of the International Coordinating
  Committee on Speech Databases and Speech I/O Systems and Assessment
  (O-COCOSDA)}.\hskip 1em plus 0.5em minus 0.4em\relax IEEE, 2017, pp. 1--5.

\bibitem{du2018aishell}
J.~Du, X.~Na, X.~Liu, and H.~Bu, ``Aishell-2: transforming mandarin asr
  research into industrial scale,'' \emph{arXiv preprint arXiv:1808.10583},
  2018.

\bibitem{yebai2018clmad}
Y.~Bai, J.~Tao, J.~Yi, Z.~Wen, and C.~Fan, ``{CLMAD}: A chinese language model
  adaptation dataset,'' in \emph{The Eleventh International Symposium on
  Chinese Spoken Language Processing (ISCSLP 2018)}, 2018.

\bibitem{li2007scalable}
J.~Li and M.~Sun, ``Scalable term selection for text categorization,'' in
  \emph{Proceedings of the 2007 Joint Conference on Empirical Methods in
  Natural Language Processing and Computational Natural Language Learning
  (EMNLP-CoNLL)}, 2007.

\bibitem{rousseau2013xenc}
A.~Rousseau, ``Xenc: An open-source tool for data selection in natural language
  processing,'' \emph{The Prague Bulletin of Mathematical Linguistics}, vol.
  100, pp. 73--82, 2013.

\bibitem{moore2010intelligent}
R.~C. Moore and W.~Lewis, ``Intelligent selection of language model training
  data,'' in \emph{Proceedings of the ACL 2010 conference short papers}.\hskip
  1em plus 0.5em minus 0.4em\relax Association for Computational Linguistics,
  2010, pp. 220--224.

\bibitem{park2019specaugment}
D.~S. Park, W.~Chan, Y.~Zhang, C.-C. Chiu, B.~Zoph, E.~D. Cubuk, and Q.~V. Le,
  ``{SpecAugment: A Simple Data Augmentation Method for Automatic Speech
  Recognition},'' \emph{Proc. Interspeech 2019}, pp. 2613--2617, 2019.

\bibitem{kingma2014adam}
D.~P. Kingma and J.~Ba, ``Adam: {A} method for stochastic optimization,''
  \emph{3rd International Conference on Learning Representations, {ICLR} 2015,
  San Diego, CA, USA, May 7-9, 2015, Conference Track Proceedings}, 2015.

\bibitem{ott2018scaling}
M.~Ott, S.~Edunov, D.~Grangier, and M.~Auli, ``Scaling neural machine
  translation,'' in \emph{Proceedings of the Third Conference on Machine
  Translation: Research Papers}, 2018, pp. 1--9.

\bibitem{irie2019language}
K.~Irie, A.~Zeyer, R.~Schl{\"u}ter, and H.~Ney, ``Language modeling with deep
  transformers,'' \emph{Proc. Interspeech 2019}, pp. 3905--3909, 2019.

\bibitem{sun2019adversarial}
S.~Sun, P.~Guo, L.~Xie, and M.-Y. Hwang, ``Adversarial regularization for
  attention based end-to-end robust speech recognition,'' \emph{IEEE/ACM
  Transactions on Audio, Speech, and Language Processing}, vol.~27, no.~11, pp.
  1826--1838, 2019.

\bibitem{karita2019comparative}
S.~Karita, X.~Wang, S.~Watanabe, T.~Yoshimura, W.~Zhang, N.~Chen, T.~Hayashi,
  T.~Hori, H.~Inaguma, Z.~Jiang, M.~Someki, N.~E.~Y. Soplin, and R.~Yamamoto,
  ``A comparative study on transformer vs {RNN} in speech applications,''
  \emph{{IEEE} Automatic Speech Recognition and Understanding Workshop, {ASRU}
  2019, Singapore, December 14-18, 2019}, pp. 449--456, 2019.

\bibitem{guo2020recent}
P.~Guo, F.~Boyer, X.~Chang, T.~Hayashi, Y.~Higuchi, H.~Inaguma, N.~Kamo, C.~Li,
  D.~Garcia-Romero, J.~Shi, J.~Shi, S.~Watanabe, K.~Wei, Z.~Wangyou, and
  Z.~Yuekai, ``Recent developments on espnet toolkit boosted by conformer,''
  \emph{arXiv preprint arXiv:2010.13956}, 2020.

\bibitem{Gulati2020}
\BIBentryALTinterwordspacing
A.~Gulati, J.~Qin, C.-C. Chiu, N.~Parmar, Y.~Zhang, J.~Yu, W.~Han, S.~Wang,
  Z.~Zhang, Y.~Wu, and R.~Pang, ``{Conformer: Convolution-augmented Transformer
  for Speech Recognition},'' in \emph{Proc. Interspeech 2020}, 2020, pp.
  5036--5040. [Online]. Available:
  \url{http://dx.doi.org/10.21437/Interspeech.2020-3015}
\BIBentrySTDinterwordspacing

\bibitem{fan2019unsupervised}
Z.~Fan, S.~Zhou, and B.~Xu, ``Unsupervised pre-traing for sequence to sequence
  speech recognition,'' \emph{arXiv preprint arXiv:1910.12418}, 2019.

\bibitem{an2019cat}
K.~An, H.~Xiang, and Z.~Ou, ``Cat: Crf-based asr toolkit,'' \emph{arXiv
  preprint arXiv:1911.08747}, 2019.

\bibitem{chen2017future}
X.~{Chen}, X.~{Liu}, Y.~{Wang}, A.~{Ragni}, J.~H.~M. {Wong}, and M.~J.~F.
  {Gales}, ``Exploiting future word contexts in neural network language models
  for speech recognition,'' \emph{IEEE/ACM Transactions on Audio, Speech, and
  Language Processing}, vol.~27, no.~9, pp. 1444--1454, 2019.

\bibitem{mun2018learning}
\BIBentryALTinterwordspacing
J.~Mun, K.~Lee, J.~Shin, and B.~Han, ``Learning to specialize with knowledge
  distillation for visual question answering,'' in \emph{Advances in Neural
  Information Processing Systems}, S.~Bengio, H.~Wallach, H.~Larochelle,
  K.~Grauman, N.~Cesa-Bianchi, and R.~Garnett, Eds., vol.~31.\hskip 1em plus
  0.5em minus 0.4em\relax Curran Associates, Inc., 2018, pp. 8081--8091.
  [Online]. Available:
  \url{https://proceedings.neurips.cc/paper/2018/file/0f2818101a7ac4b96ceeba38de4b934c-Paper.pdf}
\BIBentrySTDinterwordspacing

\end{thebibliography}


\begin{bibliographystyle}{IEEEtran}

\begin{bibliography}{refs}
	
\end{bibliography}
\end{bibliographystyle}

\end{document}